\documentclass[letterpaper,twocolumn,english,aps,prb,supescriptadress]{revtex4}
\usepackage[T1]{fontenc}
\usepackage[latin9]{inputenc}
\usepackage{color}
\usepackage{babel}

\usepackage{verbatim}
\usepackage{units}
\usepackage{bm}
\usepackage{amsbsy}
\usepackage{amstext}
\usepackage{graphicx}
\usepackage[unicode=true, pdfusetitle,
 bookmarks=true,bookmarksnumbered=false,bookmarksopen=false,
 breaklinks=false,pdfborder={0 0 0},backref=false,colorlinks=true]
 {hyperref}

\makeatletter

\providecommand{\tabularnewline}{\\}

\@ifundefined{textcolor}{}
{%
 \definecolor{BLACK}{gray}{0}
 \definecolor{WHITE}{gray}{1}
 \definecolor{RED}{rgb}{1,0,0}
 \definecolor{GREEN}{rgb}{0,1,0}
 \definecolor{BLUE}{rgb}{0,0,1}
 \definecolor{CYAN}{cmyk}{1,0,0,0}
 \definecolor{MAGENTA}{cmyk}{0,1,0,0}
 \definecolor{YELLOW}{cmyk}{0,0,1,0}
 }

\makeatother

\begin{document}

\preprint{V1g}

\title{Evidence for a Finite-Temperature Spin Glass Transition in a Diluted
Dipolar Heisenberg Model in Three Dimensions}

\author{Pawel Stasiak}

\affiliation{Department of Physics and Astronomy, University of Waterloo, Waterloo,
Ontario, N2L-3G1 Canada}

\author{Michel J. P. Gingras}

\affiliation{Department of Physics and Astronomy, University of Waterloo, Waterloo,
Ontario, N2L-3G1 Canada}

\affiliation{Canadian Institute for Advanced Research, 180 Dundas St. W., Toronto,
Ontario, M5G 1Z8, Canada}
\begin{abstract}
By means of parallel tempering Monte Carlo simulations we find strong
evidence for a finite-temperature spin-glass transition in a system
of diluted classical Heisenberg dipoles randomly placed on the sites
of a simple cubic lattice. We perform a finite-size scaling analysis
of the spin-glass susceptibility, $\chi_{{\rm SG}}$, and spin-glass
correlation length, $\xi_{L}$. For the available system sizes that
can be successfully equilibrated, the crossing of $\xi_{L}/L$ versus
temperature is strongly affected by corrections to scaling and possibly
by a short-length scale ferromagnetic spin blocking. Similarly to
many studies of different three dimensional spin-glass systems, we
do not find a crossing of the spin glass order parameter Binder ratio. 
\end{abstract}

\date{\today }

\maketitle

\section{Introduction}

Our most formal current theoretical understanding of the spin-glass
(SG) phase is based on the replica symmetry breaking (RSB) picture
set by the Parisi solution\cite{Parisi,Mezard} of the infinite-dimensional
Sherrington-Kirkpatrick model.\cite{SK} As the upper critical dimension
(UCD) of SG models is large ($d_{{\rm UCD}}$=6),\cite{Fischer} such
mean-field description is likely to be unsuitable to understand the
physics of real materials exhibiting glassy behavior. An alternative
description of the SG phase in finite dimension is given by the phenomenological
droplet picture,\cite{FisherHuse} which has been found to possibly
characterize better three dimensional (3D) SG.\cite{Jorg} However,
it remains an open debate what is the proper theory describing SG
systems in real (finite) dimensions. Most of our knowledge about the
properties of 3D SG models has been assembled thanks to years of extensive
numerical simulations. Unfortunatelly, the slow relaxation characterizing
spin glass systems makes the numerical studies very difficult. Furthermore,
the largest fraction of the numerical work has so far concentrated
on the Edwards-Anderson (EA) model\cite{EA} of $n$-component spins
interacting via nearest-neighbor random exchange interaction, $J_{ij}$,
where both ferromagnetic or antiferromagnetic couplings are present.
The cases $n$=1 and $n$=3 refer to Ising and Heisenberg SG, respectively.
The probability distribution of the random bonds, $P(J_{ij})$, is
usually taken to be Gaussian or bimodal.\cite{Binder_RevModPhys,Fischer}

A great part of the numerical studies of SG models has been devoted
to the minimal EA model, the one-component Ising SG. Due to severe
technical difficulties, only very limited range of system sizes was
accessible in the early simulations, while scaling corrections in
SG systems are large. As a result, the existence of a finite temperature
SG transition in 3D Ising SG model had remained under debate for a
long time.\cite{Ogielski_Morgenstern_Ising_SG,Bhatt_Young_Ising_SG,Ballesteros_Ising_SG,katzgraber:224432,Hasenbusch_Ising_SG,See_Hasenbusch_Ising_SG}
The early MC studies strongly supported the finite-temperature SG
transition, but zero-temperature transition could not be definitely
ruled out.\cite{Ogielski_Morgenstern_Ising_SG,Bhatt_Young_Ising_SG,Marinari}
Only quite recently, in the course of large-scale Monte Carlo studies,
has the existence of a thermodynamic phase transition in the Ising
case been seemingly firmly established\cite{Ballesteros_Ising_SG}
and, perhaps most satisfactorily, universality among systems with
different bond distributions been confirmed.\cite{katzgraber:224432,Hasenbusch_Ising_SG}

The case of the Heisenberg SG still remains somewhat more controversial
than of the Ising SG. Originally, it was believed that the lower critical
dimension (LCD) for the Heisenberg SG is $d_{{\rm LCD}}\geq3$,\cite{Morris}
and that the small anisotropies present in the real system are responsible
for the SG behavior observed in experiments.\cite{Morris,See_Gingras_PRL}
While compelling, this suggestion has some difficulties since no crossover
from Heisenberg to Ising SG universality class caused by a weak anisotropy
has ever been observed in experiments.\cite{Kawamura1487,KawamuraChirality}
It was suggested that in the Heisenberg EA SG, a finite-temperature
transition occurs in the chiral sector\cite{KawamuraChirality,Kawamura1487}
while a SG transition in the spin sector occurs at zero temperature.\cite{Kawamura1998,Hukushima_Kawamura_2005,KawamuraChirality}
The chirality is a multi-spin variable representing the handedness
of the noncolinear or noncoplanar spin structures.\cite{KawamuraChirality,Kawamura1487}
It has been proposed that the SG phase in Heisenberg SG materials
is caused by a spin-chirality coupling induced by small anisotropies.\cite{KawamuraChirality}
Later simulations indicated the existence of a nonzero-temperature
SG transition.\cite{Campos,Lee_Young} The most recent work on the
3D Heisenberg SG suggests that the SG transition may be decoupled
from the chiral glass (CG) transition, occurring at slightly higher
temperature,\cite{Viet_Kawamura,Viet_Kawamura_arXiv} or that a common
transition temperature may exist.\cite{Fernandez_Young} In two dimensions,
recent defect-wall renormalization group calculations suggest that
both the SG and CG transitions occur at zero temperature, but that
the two transitions are decoupled.\cite{Weigel_PRL,Weigel_PRB,Weigel_JPC}

On the experimental side there exist some SG materials with a strong
Ising-like uniaxial anisotropy, but the majority of experimental SG
studies focus on nearly isotropic, Heisenberg-like systems. A well
studied Ising SG material is Fe$_{0.5}$Mn$_{0.5}$TiO$_{3}$,\cite{Ito_FeMnTiO3,Gunnarsson_FeMnTiO3}
as other examples of an Ising SG more recently found, Eu$_{0.5}$Ba$_{0.5}$MnO$_{3}$\cite{Nair_EuBaMnO}
and Cu$_{0.5}$Co$_{0.5}$Cl$_{2}$-FeCl$_{3}$ \cite{Suzuki_CuCoClFeCl,Suzuki_CuCoClFeCl_2008}
can be mentioned. Considering Fe$_{0.5}$Mn$_{0.5}$TiO$_{3}$,\cite{Ito_FeMnTiO3,Gunnarsson_FeMnTiO3}
the leading coupling is a nearest-neighbor exchange, and the compounds,
FeTiO$_{3}$ and MnTiO$_{3}$, are antiferromagnets. In both cases,
the nearest-neighbor exchange interactions within the hexagonal layers
is antiferromagnetic. The magnitude of the intralayer coupling is
substantially larger than the interlayer coupling. The SG nature of
the mixture, Fe$_{0.5}$Mn$_{0.5}$TiO$_{3}$, originates from the
fact that the coupling between the layers is ferromagnetic in FeTiO$_{3}$
but antiferromagnetic in MnTiO$_{3}$;\cite{Ito_FeMnTiO3} hence,
in the mixture, random frustration occurs. For the Heisenberg case,
some short-range SG compounds are also available, for example insulating
Eu$_{x}$Sr$_{1-x}$S.\cite{Fischer,Maletta_EuSrS_SG} In the Eu-rich
case, this material is a ferromagnet; the nearest-neighbor exchange
interaction between Eu$^{2+}$ ions is ferromagnetic and the next-nearest-neighbor
exchange is weaker and antiferromagnetic.\cite{Bohn_EuS,Fischer}
When magnetic Eu is randomly substituted with nonmagnetic Sr, a random
frustration of the ferromagnetic and antiferromagnetic bonds arises.
But the most often studied SG are nearly isotropic (Heisenberg) metallic
systems interacting via a long-range Ruderman-Kittel-Kasuya-Yoshida
(RKKY) interaction between localized magnetic moments mediated by
conduction electrons. In this category, the classical systems are
alloys of noble metals such as Ag, Au, Cu or Pt, doped with a transition
metal, such as Fe or Mn, often labeled as canonical SGs. In the large
$r$ limit, where $r$ is the distance separating the magnetic moments,
the RKKY interaction varies with $r$ as $\cos(2k_{{\rm F}}r)/r^{3}$,
where $k_{{\rm F}}$ is the Fermi wavevector.

An another class of SG materials consists of spatially disordered
magnetic dipoles. The dipolar interaction has either ferromagnetic
or antiferromagnetic character depending on the relative position
of the interacting dipoles. In the presence of positional disorder,
this gives rise to random frustration and a SG phase at low temperature
and sufficiently high level of disorder is expected.\cite{foot_Ayton}\nocite{Ayton_PRL, Ayton_PRE}
A number of dipolar Ising SG materials have been identified and related
models have been studied numerically. With the aim of modeling nanosized
magnetic particles dispersed in a frozen nonmagnetic solvent,\cite{Luo_frozen_ferrofluid}
systems of Ising dipoles on fully occupied\cite{Fernandez:064404}
and diluted\cite{Fernandez_dilute} (with $x$=35\% and $x$=50\%
occupancy) simple cubic (SC) lattice with randomly oriented easy axes
have been simulated. In three dimensions, a spin-glass transition
has been identified, both in the diluted\cite{Fernandez_dilute} and
undiluted case.\cite{Fernandez:064404} A well known physical realization
of a diluted dipolar Ising model and dipolar SG is LiHo$_{x}$Y$_{1-x}$F$_{4}$.\cite{Reich_LiHo,Ancona_LiHo,Quilliam_LiHo,Tam_LiHo}
Early on, some authors suggested the existence of an exotic anti-glass
phase at very low concentration, $x$,\cite{Reich_LiHo,Ghosh-Nature,Reich_LiHo_antiglassPRL}
or questioned the existence of SG transition in LiHo$_{x}$Y$_{1-x}$F$_{4}$
altogether.\cite{Jonsson_LiHo,Jonsson-new} As well, early numerical
studies of diluted Ising dipoles on SC lattice\cite{SniderYu:214203}
and for a lattice geometry corresponding to LiHo$_{x}$Y$_{1-x}$F$_{4}$\cite{Biltmo:054437}
did not find a spin-glass transition in diluted dipolar Ising systems.\cite{SniderYu:214203,Biltmo:054437}
A more recent work, however, reports a spin-glass phase in a model
approximating LiHo$_{x}$Y$_{1-x}$F$_{4}$.\cite{Tam_LiHo} As in
previous work,\cite{Biltmo:054437} crossing of the spin-glass Binder
ratio plots was not found in Ref.~[\onlinecite{Tam_LiHo}], but a
finite-size scaling of the spin-glass correlation length provided
a compelling evidence for a thermodynamical phase transition.\cite{Tam_LiHo}
Apparently, the corrections to scaling are large for the sizes of
dipolar systems studied, being very pronounced in the Binder ratio
while the SG correlation length is somewhat less affected.\cite{Tam_LiHo}
Very recent MC simulations of a site-diluted SC lattice of Ising spins
coupled via a long-range dipolar interactions also found a finite-temperature
spin-glass transition, but with different value of the correlation
length exponent $\nu$=0.95,\cite{Alonso_arXiv} compared with $\nu$=1.3
reported in Ref.~[\onlinecite{Tam_LiHo}].

The case of dipolar Heisenberg SG is an obvious extension of the SG
phenomenology reviewed above. In the presence of spatial disorder,
the off-diagonal terms in the dipolar interaction destroy the rotational
symmetry of the ground state. Thus, the dipolar Heisenberg SG is expected
to be in the Ising universality class.\cite{Bray_Moore,See_Gingras_PRL,Bray_Moore_PRL}
In this context, it would be interesting to study a three component
($n$=3, Heisenberg) SG system where anisotropic long-range interactions,
i.e. dipolar interactions, dominate. Finding in such system critical
exponents that are consistent with the exponents of Ising SG universality
class would further confirm universality in spin glasses and boost
our confidence in our largely numerically-based understanding of real
spin glass systems.

Experimentally, a diluted dipolar SG can be realized by sufficiently
diluting magnetic dipoles with a nonmagnetic substituent, to the point
that a short-range exchange interaction becomes insignificant, and
long-range dipolar interaction dominates. The best candidate materials
are compounds containing rare earth magnetic ions, as due to the screening
of the partially filled 4$f$ shell by outer shells, the exchange
interaction is relatively weak among rare earths, while their magnetic
moments can be large. 

It was mentioned above that Eu$_{x}$Sr$_{1-x}$S, at concentration
$x\simeq0.5$, is an example of a short-range Heisenberg SG, where
the SG freezing is driven by the frustrated nearest-neighbor and next-nearest-neighbor
exchange interactions. The 4$f$ electrons of the rare earth Eu$^{2+}$
ion give an $^{8}S_{7/2}$ ground state with a sizable magnetic moment
of 7$\mu_{{\rm B}}$.\cite{Bohn_EuS} Below the percolation threshold,
i.e. $x_{c}=0.136$ for the face centered cubic lattice (FCC) with
first- and second-nearest-neighbor interactions,\cite{Dalton_Xc}
the existence of a dipolar SG in Eu$_{x}$Sr$_{1-x}$S was suggested.\cite{Tholence_EuSrS_dSG}
To explain the two maxima in the ac susceptibility, at temperatures
of order of 100 mK and 10 mK, an interplay of dipolar freezing and
blocking of small clusters was proposed.\cite{Tholence_EuSrS_dSG}
But the authors of subsequent studies\cite{Kotzler_EuSrS_dSG} suggested
that the features in the ac susceptibility of Eu$_{x}$Sr$_{1-x}$S
at concentration $0.05<x<0.13$ should be interpreted as a spin blocking
and not a dipolar SG freezing. Nevertheless, it was also proposed
that maybe at lower concentration, $x$, a dipolar SG freezing in
this material could be studied.\cite{Kotzler_EuSrS_dSG} It should
be reminded that the existence of a dipolar SG at high dilution in
LiHo$_{x}$Y$_{1-x}$F$_{4}$ has also been much questioned over the
years.\cite{Reich_LiHo,Ancona_LiHo,Quilliam_LiHo,Jonsson_LiHo,Jonsson-new}
Notwithstanding that recent experimental work reports a SG transition
in LiHo$_{x}$Y$_{1-x}$F$_{4}$($x$=0.045),\cite{Quilliam_LiHo}
it is interesting to ask if difficulties in establishing the existence
of a dipolar SG in strongly diluted Eu$_{x}$Sr$_{1-x}$S\,\cite{Tholence_EuSrS_dSG,Kotzler_EuSrS_dSG}
and in LiHo$_{x}$Y$_{1-x}$F$_{4}$, i.e. the suggested anti-glass
phase at $x$=0.045\,\cite{Reich_LiHo,Ghosh-Nature,Reich_LiHo_antiglassPRL}
or absence of signatures of SG transition both at $x$=0.045 and $x$=0.165,\cite{Jonsson_LiHo,Jonsson-new}
are related. A clearer picture of the formation of relatively large
spin blocks below the percolation threshold that collectively form
a cluster SG phase was obtained from experiments on Eu$_{x}$Ca$_{1-x}$B$_{6}$.\cite{Wigger_EuCaB}
Clusters with magnetic moments $\mu\simeq260\mu_{{\rm B}}$ were observed
and the transition temperature separating the cluster glass and paramagnetic
phases, for Eu concentrations between around $x$=0.1 and $x$=0.3,
was measured to be of order of 2 K.\cite{Wigger_EuCaB}

Promising candidates for diluted dipolar Heisenberg SGs can be found
among gadolinium compounds. Gd$^{3+}$ ion has a half-filled 4$f$-shell.
The ground state manifold is $^{8}S{}_{7/2}$ and with little orbital
momentum contribution. Gd$^{3+}$ is therfore a good approximation
of a classical Heisenberg spin. Good example of materials that can
be considered as candidates for diluted dipolar Heisenberg SGs are
(Gd$_{x}$Y$_{1-x}$)$_{2}$Ti$_{2}$O$_{7}$ and (Gd$_{x}$Y$_{1-x}$)$_{2}$Sn$_{2}$O$_{7}$.
Gd$_{2}$Ti$_{2}$O$_{7}$ and Gd$_{2}$Sn$_{2}$O$_{7}$ are strongly
frustrated Heisenberg pyrochlore antiferromagnets.\cite{Raju_GdTi,Bondah_GdSn,Gardner_RMP}
While the Curie-Weiss temperature is about $\theta_{{\rm CW}}\sim-10$
K, due to frustration, both compounds remain disordered down to $T=$1~K.\cite{Raju_GdTi,Bonville_GdSn_GdTi}
Theoretically, the extensive ground-state degeneracy in the pyrochlore
nearest-neighbor Heisenberg antiferromagnet prevents ordering down
to zero temperature,\cite{Moessner_Chalker_cl_sl,Canals_qn_sl} and
the low temperature order in the aforementioned materials is induced
by other, weaker interactions that are specific to each of these compounds.
One of the interaction at play below 1 K is the dipolar interaction.
Indeed, in the case of Gd$_{2}$Sn$_{2}$O$_{7}$, the spin configuration
in the ordered state was found\cite{Wills_GdSn} to be the ground
state of the pyrochlore antiferromagnet with dipolar interactions.\cite{Palmer_Chalker}
In the case of Gd$_{2}$Ti$_{2}$O$_{7}$, other interactions, like
further nearest-neighbor exchange, are likely at play. Indeed, below
the first phase transition at 1 K, there is another one at 0.7 K,\cite{Bonville_GdSn_GdTi}
with both phases ordered with propagation vector $k=\left[\frac{1}{2}\,\frac{1}{2}\,\frac{1}{2}\right]$.\cite{Champion_GdTi,Stewart_GdTi}

Another candidate material, also well studied in the context of geometrical
frustration, is the Gd$_{3}$Ga$_{5}$O$_{12}$ garnet (GGG). The
cubic lattice structure of this frustrated Heisenberg antiferromagnet
consist of two interpenetrating sublattices of corner-sharing triangles.
While the Curie-Weiss temperature is $\theta_{{\rm CW}}\sim$-2 K,
the frustration postpones ordering down to 0.18 K.\cite{Petrenko_GGG}
The rich low temperature physics of GGG is still not fully understood,
but some insight has been recently gained from dynamic magnetization
studies, revealing that in the low temperature phase there is long-range
order coexisting with both spin liquid\cite{Dunsiger_GGG_sl,Ghosh_GGG}
and spin glass\cite{Schiffer_GGG_sg} behavior. In the framework of
Gaussian mean-field theory, it was shown that the dipolar interaction
plays an important role in the ordering in GGG,\cite{Yavors'kii_GGG,Yavors'kii_GGG_2}
and that the neutron scattering data\cite{Petrenko_GGG} can be reproduced
with a proper treatment of the dipolar interactions.\cite{Yavors'kii_GGG,Yavors'kii_GGG_2}
Analogously to (Gd$_{x}$Y$_{1-x}$)$_{2}$Ti$_{2}$O$_{7}$ and (Gd$_{x}$Y$_{1-x}$)$_{2}$Sn$_{2}$O$_{7}$,
at sufficient dilution, (Gd$_{x}$Y$_{1-x}$)$_{3}$Ga$_{5}$O$_{12}$
may be expected to exhibit, at low temperature, a dipolar SG phase. 

In the studies motivated by experiments indicating a SG freezing in
frozen ferrofluids,\cite{Luo_frozen_ferrofluid} some evidence for
a SG freezing in a system of dense amorphous Heisenberg and XY spins
coupled by long-range dipolar interactions was obtained from molecular
dynamics simulations.\cite{Ayton_PRL,Ayton_PRE} But no systematic
investigation of the thermodynamic nature of the freezing was at that
time really possible.

In the present work, in anticipation of eventual experimental studies
of dipolar SG, e.g. diluted Gd compounds, or further work on Eu$_{x}$Sr$_{1-x}$S
or Eu$_{x}$Ca$_{1-x}$B$_{6}$, we perform numerical studies of the
SG transition in a diluted dipolar Heisenberg model. At high dilution,
the lattice structure should be irrelevant and data obtained for different
systems should be comparable.\cite{foot_ignore} Here we consider
the simplest possible geometry where we study dipoles randomly placed
at the sites of SC cubic lattice. We provide Monte Carlo data that
supports the scenario that, at low dipole concentration, the diluted
dipolar Heisenberg model displays an equilibrium phase transition
to a SG phase. We calculate the critical exponents $\nu$ and $\eta$
for the SG transition in the model studied. The derived exponents
do not match experimental or Monte Carlo exponents, neither for Heisenberg
nor Ising SG. This may be because of important scaling corrections
and severe restriction of the system sizes that we were able to study
due to the computationally expensive summation of long-range dipole-dipole
interaction and very slow equilibration.

The rest of the paper is organized as follows. In Section~\ref{sec:Model-and-method},
we define the model and MC method employed. In Section~\ref{sec:Physical-Quantities},
we introduce the observables calculated in the simulation. In Section~\ref{sec:Monte-Carlo-Results},
we present and discuss our results. Our conclusions are in Section~\ref{sec:Summary}.
Some technical details are discussed in Appendices. In Appendix \ref{sec:Magnetization},
we examine the temperature and system size dependence of the magnetization
and staggered magnetization of the model studied. In Appendix~\ref{sec:Self-interaction},
we discuss the issue of the self-interaction term that must be taken
into account when periodic boundary conditions are imposed. In Appendix~\ref{sec:Ewald},
we discuss the Ewald summation technique. In Appendices~\ref{sec:Overrelaxation}
and \ref{sec:Heatbath-algorithm}, we discuss the overrelaxation and
heatbath algorithms, respectively.

\section{Model and method\label{sec:Model-and-method}}

\subsection{Model}

We consider a system that consists of classical three-component ($n$=3,
Heisenberg) dipoles that are free to point in any direction. The dipoles
are randomly distributed on the sites of a 3D simple cubic (SC) lattice.
The Hamiltonian is of the form\begin{equation}
\mathcal{H}=\frac{1}{2}\epsilon_{d}\sum_{i\neq j}\sum_{\mu,\nu}\frac{\delta^{\mu\nu}r_{ij}^{2}-3r_{ij}^{\mu}r_{ij}^{\nu}}{r_{ij}^{5}}S^{\mu}(\bm{r}_{i})S^{\nu}(\bm{r}_{j}),\label{eq:H}\end{equation}
where $S^{\mu}(\bm{r}_{i})$ ($\mu=$ $x$, $y$, $z$) denote Cartesian
components of classical spin vectors, $\bm{S}(\bm{r}_{i})$, which
are of unit length, $\left|\bm{S}(\bm{r}_{i})\right|=1$. The energy
scale of dipolar interactions is set by $\epsilon_{d}=\frac{\mu_{0}\mu^{2}}{4\pi a^{3}}$,
where $\mu$ is the magnetic moment of the spin $\bm{S}(\bm{r}_{i})$,
$a$ is the lattice constant and $\mu_{0}$ denotes vacuum permeability.
Below, $\epsilon_{d}/k_{{\rm B}}$, where $k_{{\rm B}}$ is the Boltzmann
constant, is conveniently used as a unit of temperature. The summation
is carried over all occupied lattice sites and over the vector components
of the spin, $\mu,\nu=$ $x$, $y$ and $z$. The factor $1/2$ is
included to correct for double counting of dipole pairs.%
{} $\bm{r}_{i}$ and $\bm{r}_{j}$ are the positions of ions labeled
$i$ and $j$, respectively, and their distance, $\left|\bm{r}_{ij}\right|=\left|\bm{r}_{j}-\bm{r}_{i}\right|$,
is measured in units of nearest-neighbor distance, $a$. We use periodic
boundary conditions. In the case of long-range interactions, this
means that to calculate a pairwise interaction, we must sum over an
infinite array or dipole images replicated with a periodicity set
by the dimensions of the simulation box. Therefore, it is convenient
to consider the interaction constant for spins $i$ and $j$ as a
3 by 3 matrix, $\hat{L}_{ij}$. The matrix elements of $\hat{L}_{ij}$
are denoted $L_{ij}^{\mu\nu}$, and for dipoles separated by a vector
$\bm{r}_{ij}$, are given by the sum, \begin{equation}
L_{ij}^{\mu\nu}=\sum_{\bm{n}}\frac{\delta^{\mu\nu}\left|\bm{r}_{ij}+\bm{n}\right|^{2}-3\left(\bm{r}_{ij}+\bm{n}\right)^{\mu}\left(\bm{r}_{ij}+\bm{n}\right)^{\nu}}{\left|\bm{r}_{ij}+\bm{n}\right|^{5}}.\label{eq:Lsum}\end{equation}
 Vectors $\bm{n}$ are of the form $\bm{n}=L\left(k\hat{x}+l\hat{y}+m\hat{z}\right)$,
where $k$, $l$, $m$ are integers and $\hat{x}$, $\hat{y}$ and
$\hat{z}$ are the unit vectors pointing in the directions of primitive
translation vectors of the SC lattice. $L$ is an integer expressing
the size of the simulation cell in units of the lattice constant,
$a$. Note that, in a simulation, care must be taken to correctly
include the so-called self-interaction term, $\hat{L}_{ii}$, (see
Appendix \ref{sec:Self-interaction}). The self-interaction term originates
from interaction of spin with its periodic images replicated outside
the simulation cell. The lattice summation (\ref{eq:Lsum}) is performed
using the Ewald technique;\cite{Ewald,Dove,Holm_Wang,Leeuw_Perram}
the details of which are given in Appendix \ref{sec:Ewald}. The calculated
Ewald sums correspond to a summation over a long cylinder, such that
the demagnetization field is zero. Using interaction constants defined
in Eq. (\ref{eq:Lsum}), the Hamiltonian can then be written in the
form\begin{equation}
\mathcal{H}=\frac{1}{2}\epsilon_{d}\sum_{i,j}\bm{S}(\bm{r}_{i})\hat{L}_{ij}\bm{S}(\bm{r}_{j}).\label{eq:H with L}\end{equation}
The summation in Hamiltonian (\ref{eq:H with L}) includes only the
spins enclosed in the simulation cell, while the presence of spins
outside the simulation box is approximated by periodic images of the
spins within the simulation cell, and this effect is included in the
interaction constants, given by the matrix $\hat{L}_{ij}$ calculated
via the Ewald method. Note that, unlike in Eq. (\ref{eq:H}), the
self-interaction terms, $i=j$, are included in the summation (\ref{eq:H with L}).

\subsection{Method}

Our simulations employ the standard single spin-flip Metropolis Monte
Carlo algorithm with parallel tempering (PT).\cite{MarinariParisi,Hukushima:1604}
PT was found to significantly speed up equilibration in slowly relaxing
systems.\cite{MarinariParisi,Hukushima:1604} In this technique, one
simultaneously simulates a number, $N_{T}$, of thermal replicas -
copies of the system with the same spatial disorder, but at different
temperatures. In each thermal replica, the simulation begins from
a different random initial spin configuration. At every 10 local update
sweeps, each consisting of $N$ single spin updates, where $N$ is
the number of spins in the system, a configuration swap among thermal
replicas is attempted with the acceptance probability preserving the
detailed balance condition. The frequency of tempering is chosen to
balance the following two factors. As thermal tempering is computationally
inexpensive, it is desirable to perform replica swap attempts often,
to promote traveling of the replicas along the temperature axis. But,
on the other hand, after a parallel tempering induced configuration
exchange, a sufficient number of local update sweeps has to be performed
to let the new configuration evolve at the given temperature. If a
subsequent tempering is attempted too soon, the two configurations
could be swapped back, and in that case no progress in the relaxation
of a state trapped in a local energy minimum would have been made.
The number of thermal replicas, $N_{{\rm T}}$, and simulated temperatures,
$T_{\alpha}$, where $\alpha=1,\ldots,N_{{\rm T}}$, are chosen to
yield a sufficiently high and temperature independent PT configuration
swap acceptance rate, i.e. not less than 50\%. A uniform with respect
to temperature, $T_{\alpha}$, PT acceptance rate is achieved by choosing
$T_{\alpha}$ to satisfy the formula\cite{Lee_Young}\begin{equation}
(T_{\alpha}-T_{\alpha-1})/T_{\alpha}=1/\sqrt{C_{V}\, N},\label{eq:PTdT}\end{equation}
were $N$ denotes the number of dipoles. The specific heat per spin,
$C_{V}$, used in Eq. (\ref{eq:PTdT}) was measured in preliminary
simulations of the smallest system sizes, with uniformly distributed
temperatures. 

\textcolor{black}{The Metropolis single-spin moves are attempted within
a temperature-dependent solid angle, where the angle is self-consistently
chosen such that the acceptance rate of MC single spin move is close
to 50\%. To carry out a spin move, we choose a coordinate system with
the $\hat{z}$ axis along the current spin direction, and randomly
choose a polar angle, $\theta$, and azimuthal angle, $\phi$. In
order to obtain a uniform distribution of random points on a unit
sphere one needs to draw $\phi$ and $z=\cos(\theta)$ from a uniform
probability distribution, such that $\phi\in(0,2\pi)$ and $z\in(-1,1)$.
Here, to maintain the desired acceptance rate, the move is restricted
to a limiting angle, $\theta_{{\rm max}}$, relative to the initial
spin direction; hence, the choice of $z$ is restricted to $z\in(1-z_{{\rm max}},1)$,
where $z_{{\rm max}}=\cos(\theta_{{\rm max}})$. To obtain $z_{{\rm max}}$
such that the acceptance rate, $p_{{\rm acc}}$, is 50\%, during each
100 MCS $p_{{\rm acc}}$ is measured, and afterwards $z_{{\rm max}}$
is adjusted. If $p_{{\rm acc}}$ is lower than 0.5, $z_{{\rm max}}$
should be decreased; in the opposite case, when $p_{{\rm acc}}>0.5$,
$z_{{\rm max}}$ should be increased, while for $p_{{\rm acc}}=0.5$,
$z_{{\rm max}}$ does not change. Such update of $z_{{\rm max}}$
can be obtained when multiplying the current value of $z_{{\rm max}}$,
$z_{{\rm max}}^{({\rm old})}$, by $2p_{{\rm acc}}$; hence, a new
value of $z_{{\rm max}}$ is calculated according to the formula $z_{{\rm max}}^{({\rm new})}=2p_{{\rm acc}}z_{{\rm max}}^{({\rm old})}$,
with the restriction $z_{{\rm max}}^{({\rm new})}\in(0.001,2)$. After
choosing $\phi$ and $\theta$, that is a new spin direction in the
coordinates relative to the initial spin direction, a transformation
to the global coordinate system is performed.}

We simulated two dipole concentrations, $x=1/16=0.0625$ and $x=1/8=0.125$,
and for each concentration we considered 4 system sizes varying between
around 30 and 200 dipoles, which is the largest size that we were
able to equilibrate. To perform the necessary disorder average (see
Section~\ref{sec:Physical-Quantities}), we considered at least 1000
disorder samples. The parameters of the simulations are collected
in Table~\ref{tab:Parameters}. To generate results reported here,
we used in total around $3\cdot10{}^{5}$ hours ($\sim$35 years)
of CPU time on AMD Opteron, 2.6 GHz. The statistical error is based
on disorder sampling fluctuation and is calculated using the standard
jackknife method.\cite{Quenouille_jacknife,Tukey_jacknife,Berg_book}%
\begin{table}
\begin{tabular}{|c|c|c|c|c|c|c|c|}
\hline 
$L$  & $N_{{\rm dip}}$  & $N_{{\rm samp}}$  & $N_{{\rm eq}}$  & $N_{{\rm prod}}$  & $N_{{\rm T}}$  & $T_{{\rm min}}$  & $T_{{\rm max}}$\tabularnewline
\hline
\hline 
\multicolumn{8}{|c|}{$x$=0.0625}\tabularnewline
\hline
\hline 
8  & 32  & 5000  & 5$\cdot$10$^{5}$  & 5$\cdot$10$^{5}$  & 16  & 0.05  & 0.1763\tabularnewline
\hline 
10  & 62,63  & 2000, 2000  & 2$\cdot$10$^{6}$  & 10$^{6}$  & 16  & 0.05  & 0.1763\tabularnewline
\hline 
12  & 108  & 1200  & 10$^{7}$  & 10$^{6}$  & 16  & 0.05  & 0.1763\tabularnewline
\hline 
14  & 172  & 1000  & 10$^{7}$  & 10$^{6}$  & 16  & 0.05  & 0.1763\tabularnewline
\hline
\hline 
\multicolumn{8}{|c|}{$x$=0.125}\tabularnewline
\hline
\hline 
6  & 27  & 3000  & 5$\cdot$10$^{5}$  & 5$\cdot$10$^{5}$  & 16  & 0.0750  & 0.2869\tabularnewline
\hline 
8  & 64  & 2000  & 2$\cdot$10$^{6}$  & 10$^{6}$  & 16  & 0.0750  & 0.2869\tabularnewline
\hline 
10  & 125  & 1500  & 2$\cdot$10$^{6}$  & 10$^{6}$  & 16  & 0.0800  & 0.2811\tabularnewline
\hline 
12  & 216  & 1000(+200)  & 5$\cdot$10$^{6}$(2$\cdot$10$^{7}$)  & 10$^{6}$  & 16  & 0.0850  & 0.2787\tabularnewline
\hline
\end{tabular}

\caption{Parameters of the Monte Carlo simulations for two dipole concentrations,
$x$. $L$ is the linear size of the simulation box; $N_{{\rm dip}}$
is the number of spins, and $N_{{\rm samp}}$ denotes number of disorder
samples. $N_{{\rm eq}}$ and $N_{{\rm prod}}$ are the number of MCS
in the equilibration and measurement phase of the simulation, respectively.
$N_{{\rm T}}$ is the number of thermal replica and $T_{{\rm min}}$,
$T_{{\rm max}}$ are the lowest and highest temperatures in PT scheme.
For $L$=10, $x$=0.0625, to obtain the desired $x$, two numbers of dipoles
were simulated, and the disorder average was taken over the results
for both $N_{{\rm dip}}=62$ and $N_{{\rm dip}}=63$. For $L$=12,
$x$=0.125 the numbers given in round brackets pertain to a subset
of disorder replicas simulated longer, to monitor equilibration; the
long equilibration time results for these replicas were included in
the disorder averaging.\label{tab:Parameters}}

\end{table}

To reduce the number of performed lattice sums, for each lattice site
$k$ we calculate the local interaction field, \begin{equation}
\bm{H}_{k}=\sum_{j\neq k}\hat{L}_{kj}\bm{S}_{j},\label{eq:Hk}\end{equation}
 and update it only when the spin change is accepted. Having $\bm{H}_{k}$
available, the computational complexity of calculating the energy
change when a single spin is moved, which is needed to test if the
spin move is accepted, is of order of a small constant number of arithmetic
operations, $\mathcal{O}(1)$. As the field, $\bm{H}_{k}$, is updated
only if the spin move is accepted, the computational cost of updating
the local field, $\bm{H}_{k}$, which is $\mathcal{O}(N)$, is avoided
if the spin move is rejected. In consequence the computational cost
of rejected Metropolis spin updates is negligible.

In was reported that the autocorrelation time can be substantially
decreased in simulations of Heisenberg SG by performing computationally
inexpensive overrelaxation (microcanonical) spin updates.\cite{Creutz_overrelaxation,Alonso_overrelaxation}
Overrelaxation updates are zero-energy spin moves that consist of
180 degrees rotation of the spin around the local molecular field.
Including overrelaxation moves in simulations of short-range Heisenberg
spin systems is beneficial because overrelaxation moves are faster
than Metropolis updates. Here, in the case of long-range interaction,
the computational cost of overrelaxation moves would not be less then
the cost of Metropolis spin flips, as most of the time is spent on
updating the local interaction field (\ref{eq:Hk}), and the local
interaction field has to be updated both after an accepted Metropolis
spin flip and after an overrelaxation move. Furthermore, it is worth
to note that in the general case of non-cubic geometry with periodic
boundary conditions, where the self-interac\textcolor{black}{tion
term is present (see App}endix \ref{sec:Self-interaction}\textcolor{black}{),
an overrelaxation move does not preserve the energy (see App}endix
\ref{sec:Overrelaxation}).

In the case of the nearest-neighbor Heisenberg SG model, it is more
efficient to use the heatbath algorithm\cite{Creutz_heatbath,Miyatake_heatbath,Olive_heatbath}
for local spin updates. In the heatbath algorithm spins are individually
connected to a heatbath, i.e. a new spin direction, which is independent
from the previous direction, is drawn from the Boltzmann probability
distribution for a spin in the local molecular field, ${\bm H}_{k}$. Hence,
in contrast to the conventional Metropolis method, in the heatbath
algorithm, the computational cost of rejected spin moves is avoided.
For the current problem, the benefit of using the heatbath algorithm
would not be high because the computational cost of rejected spin
update attempts is negligible in comparison with the computational
cost of accepted updates which require recalculating the lattice sums
in Eq. (\ref{eq:Hk}). Also, similarly to the case of overrelaxation
moves, if the geometry of the simulation cell is not cubic, the self-interaction
term in the Hamiltonian introduced by the periodic boundary conditions
makes the heatbath algorithm impractical (see Appendices \ref{sec:Self-interaction}
and \ref{sec:Heatbath-algorithm}). So, while we are considering here
a cubic system, in order to keep our method general and to obtain
results that are the easiest to compare with possible future simulations
with different lattice geometries, we decided to not use the heatbath
algorithm nor overrelaxation moves in this work.

\section{Physical Quantities\label{sec:Physical-Quantities}}

In spin-glass systems, the order parameter can be defined as an overlap
between two independent, identical copies of the system. In the case
of 3D Heisenberg spins, the overlap can be calculated for 9 combinations
of the vector components. We write \begin{equation}
q^{\mu\nu}(\bm{k})=\frac{1}{N}\sum_{\bm{r}}S_{\mu}^{(\alpha)}(\bm{r})S_{\nu}^{(\beta)}(\bm{r})\exp(i\boldsymbol{k}\cdot\bm{r}),\end{equation}
 where $\mu,\nu=x,y,z$ and where $\alpha$ and $\beta$ denote different
copies of the system with the same random disorder and that are simulated
simultaneously, but independently. The wave-vector-dependent SG order
parameter is \begin{equation}
q(\bm{k})=\sqrt{\sum_{\mu,\nu}\left|q^{\mu\nu}(\bm{k})\right|^{2}}.\end{equation}
 In the case of EA Heisenberg SG, in addition to spin, a chirality
variable\cite{KawamuraChirality,Kawamura1487} is also considered,\cite{Kawamura1998,Hukushima_Kawamura_2005,KawamuraChirality,Campos,Lee_Young,Viet_Kawamura,Viet_Kawamura_arXiv,Fernandez_Young}
and the chirality overlap order parameter and further quantities defined
using this order parameter are calculated. Here, for a diluted dipolar
Heisenberg SG, we do not consider chirality. The chirality cannot
be easily defined in a diluted system. Moreover, note that, as discussed
in the Introduction, the anisotropy of the dipolar interaction, in
the presence of spatial disorder, brakes the rotational symmetry.
Because of that, the chirality, if it was defined, cannot be decoupled
from the spin.

Traditionally, the finite-size scaling (FSS) analysis of SG simulation
data has been based on the calculation of Binder ratios,\cite{Binder_UL,Landau_Binder_book,Binder_UL_PRL}
which, for an $n$=3 Heisenberg SG, is defined as:\cite{Viet_Kawamura,Fernandez_Young}\begin{equation}
U_{L}=\frac{1}{2}\left(11-9\frac{\left[\left\langle q(0)^{4}\right\rangle \right]}{\left[\left\langle q(0)^{2}\right\rangle \right]^{2}}\right),\label{eq:Binder}\end{equation}
 where $\left\langle \ldots\right\rangle $ denotes thermal averaging
and $\left[\ldots\right]$ is a disorder average. The numerical factors
in Eq. (\ref{eq:Binder}) are chosen such that at $T=\infty$, assuming
Gaussian distribution of $q(0)$, $U_{L}=0$, and at $T=0$, where
$q(0)$ is not fluctuating, $U_{L}$ is 1. Being a dimensionless quantity,
$U_{L}$ is expected to display FSS properties described by\cite{katzgraber:224432}\begin{equation}
U_{L}=\tilde{X}(L^{1/\nu}(T-T_{g})),\end{equation}
where the scaling function\cite{foot_universal}\nocite{Privman_scaling}
$\tilde{X}$ is an analytic function of its argument, and $\nu$ is
the universal correlation length exponent, such that there is no system
size dependence outside the argument of the scaling function. Many
recent works report that in the case of disordered spin glass systems,
a better FSS analysis can be achieved when considering the finite-size
SG correlation length, $\xi_{L}$.\cite{Ballesteros_Ising_SG,Lee_Young,Beath:10A506,Tam_LiHo}
In the context of Ising SGs, it was suggested that $U_{L}$ may not
cross due to a lack of unique ground state\cite{Beath:10A506} or
because it is too noisy (see footnote Ref.~[\onlinecite{foot_smoothUL}]),\cite{Ballesteros_Ising_SG}
being a quantity that requires evaluation of a four-point correlation
function, as opposite to $\xi_{L}$, that is defined using a two-point
correlation function. It was also observed that scaling corrections
are larger for $U_{L}$ than for $\xi_{L}$.\cite{Ballesteros_Ising_SG}
It is likely that in the case of Heisenberg SG large scaling corrections
are the leading factor behind the lack of crossing of the Binder ratios.
To proceed, we define the SG susceptibility\cite{Binder_RevModPhys,Ballesteros_Ising_SG,Lee_Young}
as\begin{equation}
\chi_{{\rm SG}}(\bm{k})=N\left[\left\langle q(\bm{k})^{2}\right\rangle \right].\label{eq:chi}\end{equation}
 Assuming an Ornstein-Zernike form for the SG susceptibility,\cite{Schwabl_Ornstein-Zernike_book}\begin{equation}
\chi_{{\rm SG}}(\bm{k})\propto1/(\left|\bm{k}\right|^{2}+\xi^{-2}),\end{equation}
 where $\left|\bm{k}\right|\ll1/\xi$. We define a finite-size SG
correlation length,\cite{Cooper_xi,Ballesteros_Ising_SG} $\xi_{L}$,
via\begin{equation}
\xi_{L}=\frac{1}{2\sin(k_{\mathrm{min}}/2)}\left(\frac{\chi_{{\rm SG}}(0)}{\chi_{{\rm SG}}({\bf k}_{\mathrm{min}})}-1\right)^{1/2}.\end{equation}
 The correlation length divided by the system dimension, $\xi_{L}/L$,
similarly to the Binder ratio, is a dimensionless quantity that is
expected to scale according to the relation\cite{Ballesteros_Ising_SG,katzgraber:224432,Fernandez_Young,Lee_Young}\begin{equation}
\xi_{L}/L=\tilde{Y}(L^{1/\nu}(T-T_{g})),\label{eq:xi scaling}\end{equation}
 where $\tilde{Y}$ is once again a scaling function. Hence, at a
putative SG transition temperature, $T_{g}$, $\xi_{L}/L$ is expected
to be size independent.

\section{Monte Carlo Results\label{sec:Monte-Carlo-Results}}

A system of Heisenberg dipoles on a fully occupied SC lattice orders
antiferromagnetically.\cite{Luttinger_Tisza,Romano} To rule out a
long-range order in the simulated diluted systems, we calculated the
magnetization, $M$, and the staggered magnetization, $M_{{\rm stag}}$.
Both $M$ and $M_{{\rm stag}}$ are small and decrease with increasing
system size. This indicates that their nonzero value is a finite-size
effect and not a result of long-range ordering. More detailed discussion
of $M$ and $M_{{\rm stag}}$ is given in Appendix~\ref{sec:Magnetization}.
\begin{figure}[h]
 \includegraphics[width=0.5\columnwidth]{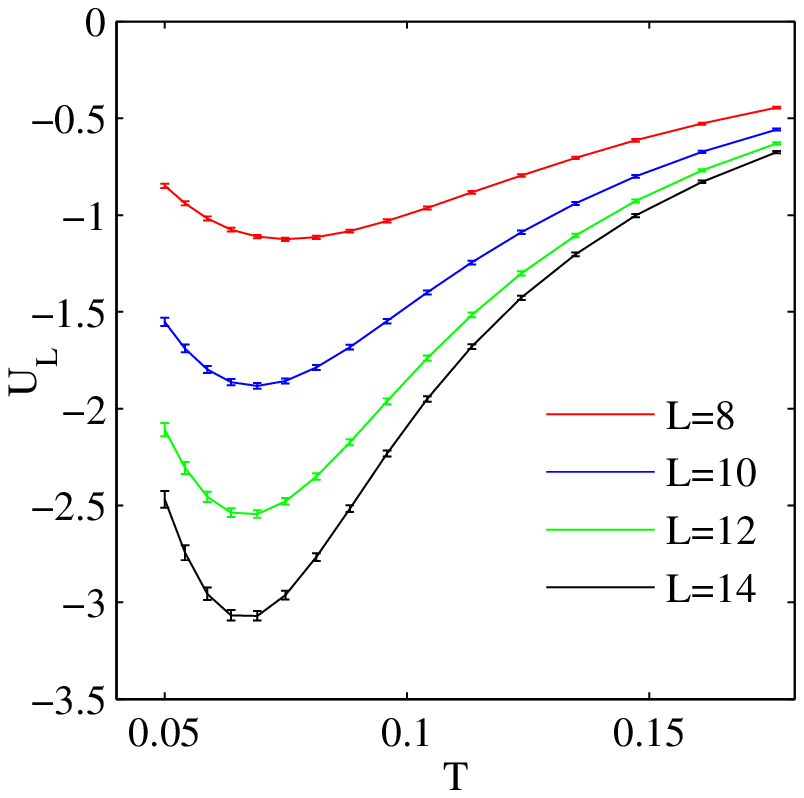}\includegraphics[width=0.5\columnwidth]{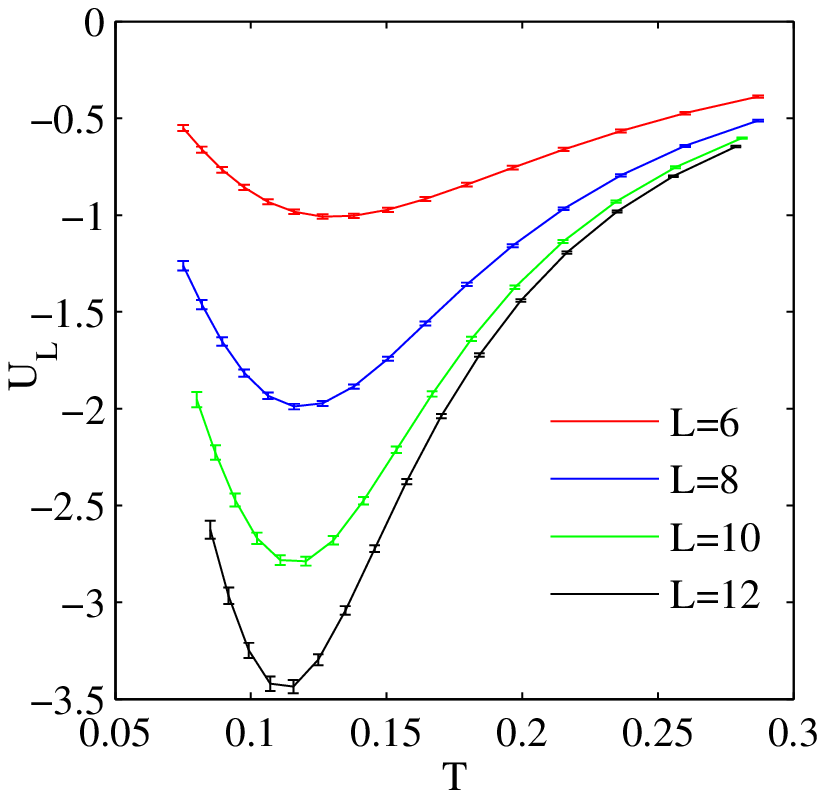}

\caption{(color online).
Binder ratios for $x$=0.0625 (left) and $x$=0.125 (right) as a function
of temperature.\label{fig:UL625_125}}

\end{figure}

We plot in Fig. \ref{fig:UL625_125} the temperature dependence of
the Binder ratio, $U_{L}$, for $x$=0.125 and $x$=0.0625, for different
system sizes. The Binder ratio curves do not cross; hence, they do
not provide indication of a phase transition. Also, in some studies
of other models, a crossing of the Binder ratios was not found, while
the scaling invariance of the finite-size correlation length was established,
indicating a transition to a SG phase. The magnitude of scaling corrections
is different for different observables and they are likely to be larger
for Binder ratio than for correlation length. In the simulation of
the Ising EA SG\cite{Ballesteros_Ising_SG,katzgraber:224432} $U_{L}$
does cross, but the scaling corrections are found to be larger for
$U_{L}$ than for $\xi_{L}/L$.\cite{Ballesteros_Ising_SG} In the
case of the site diluted EA Ising SG,\cite{Jorg_dil_EA} where scaling
corrections are large in comparison with other Ising SG models, $\xi_{L}/L$
plots are crossing with large shifts between system sizes, while the
$U_{L}$ curves do not cross, but merge at low temperature. A similar
effect has been seen in the studies of diluted dipolar Ising SG\cite{Tam_LiHo,Biltmo:054437}
- $U_{L}$ plots do not cross, but they have a tendency to merge at
low $T$, while $\xi_{L}/L$ plots intersect.\cite{Tam_LiHo} In the
case of isotropic Heisenberg EA SG,\cite{Viet_Kawamura,Viet_Kawamura_arXiv,Fernandez_Young}
the behavior of spin and chirality Binder ratios differs, but neither
show a crossing, while the correlation length shifts between system
sizes shows that scaling corrections are large. It is worthwhile to
note that the form of Binder ratio plots, characterized by a dip to
a negative value in the proximity of $T_{g}$, resembles the Binder
ratio plots for chirality (and not spin) in the Heisenberg EA model,\cite{Viet_Kawamura,Fernandez_Young}
or the Binder ratio plots for spin in Heisenberg SG models in the
presence of random anisotropy in three\cite{Matsubara_ranis_3D} and
four\cite{Shirakura_ranis_4D} dimensions.

\begin{figure}[h]
\includegraphics[width=0.8\columnwidth]{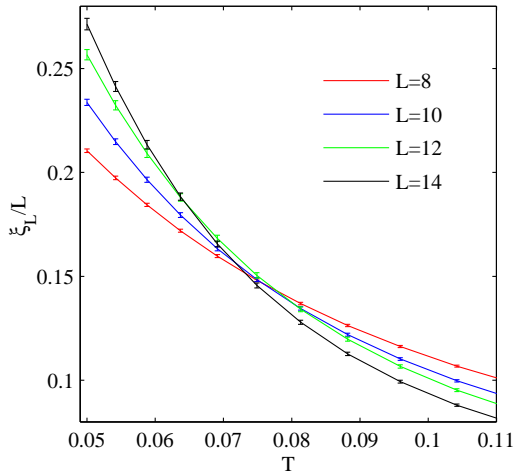}

\caption{
(color online).
SG correlation length as a function of temperature, $x$=0.0625. \label{fig:xi625}}

\end{figure}

\begin{figure}[h]
\includegraphics[width=0.8\columnwidth]{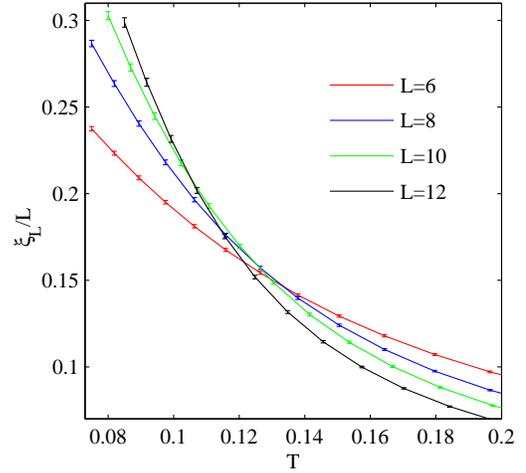}

\caption{(color online).
SG correlation length as a function of temperature, $x$=0.125. \label{fig:xi125}}

\end{figure}

Having discussed the temperature dependence of the Binder ratios,
we now turn to the behavior of the SG correlation length, $\xi_{L}(T)$.
We show the plots of $\xi_{L}/L$ vs $T$ for various system sizes
in Figs. \ref{fig:xi625} and \ref{fig:xi125}. The curves do cross;
but, for both concentrations, there are large shifts between the intersection
points for different system sizes. The large shifts between the intersection
points were also found in the studies of the EA Heisenberg SG.\cite{Viet_Kawamura,Viet_Kawamura_arXiv,Fernandez_Young}
In the case of EA Heisenberg SG, a broad range of system sizes were
studied and the shifts between the intersection points were systematically
analyzed.\cite{Viet_Kawamura,Viet_Kawamura_arXiv,Fernandez_Young}
The limitation with our data, which are due to time consuming summation
of long range interactions, prevent us from investigating large system
sizes and, therefore, to perform such an analysis. Because of a narrow
range of available system sizes, the separation between the curves
in the crossing region is small in comparison with the errorbars. Hence,
the statistical uncertainty of locating the crossing points would
be large. Indeed, looking at Figs. \ref{fig:xi625} and \ref{fig:xi125}
one realizes that by moving the curves within the error bars, the
position of the crossing can be changed substantially. Also, the number
of system sizes and, consequently, the number of intersection points
is small. In our $\xi_{L}/L$ vs $T$ data, the shifts between the
intersection points for the smallest system sizes, consistently for
both concentrations, are much smaller than the shift of the intersection
point of the two largest system sizes studied. Such feature have not
been found in simulations of EA Heisenberg SG.\cite{Viet_Kawamura,Viet_Kawamura_arXiv,Fernandez_Young}
A possible explanation for such behavior may be existence of short-range
ferromagnetic correlations. Such a ferromagnetic spin blocking would
especially strongly affect the data for the smallest system sizes,
which possibly have the linear dimensions comparable with the length
scale of the short-range ferromagnetic correlations. Spin blocking
has been observed experimentally in studies of diluted dipolar Heisenberg
systems Eu$_{x}$Sr$_{1-x}$S~\cite{Tholence_EuSrS_dSG,Kotzler_EuSrS_dSG}
and Eu$_{x}$Ca$_{1-x}$B$_{6}$.\cite{Wigger_EuCaB} Another argument
supporting short range ferromagnetic correlations scenario in the
model studied herein is the finite-size magnetization found for small
system sizes (see Appendix~\ref{sec:Magnetization}). 

\begin{figure}[h]
\includegraphics[clip,width=0.8\columnwidth]{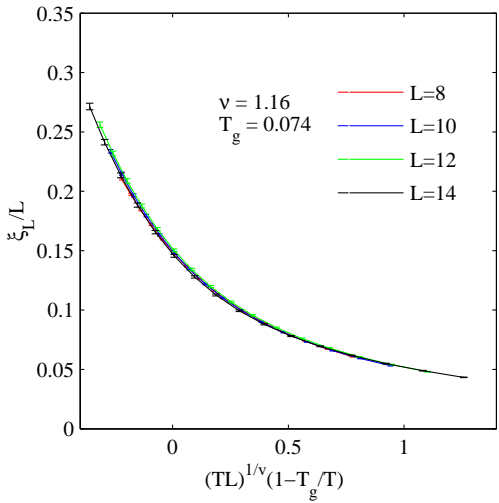}

\caption{(color online).
Extended scaling of $\xi_{L}/L$ at $x$=0.0625\label{fig:sc625}}

\end{figure}

\begin{figure}[h]
 \includegraphics[width=0.8\columnwidth]{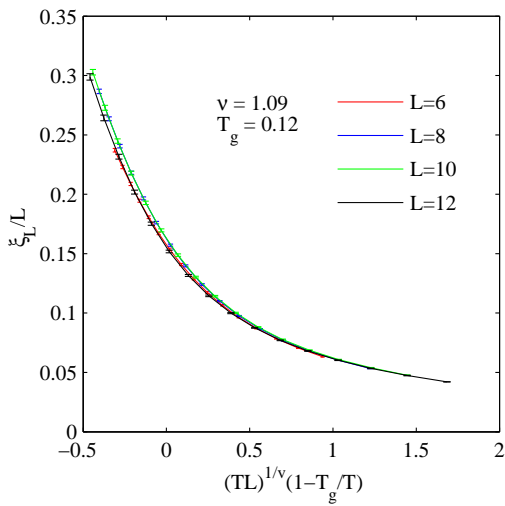}

\caption{(color online).
Extended scaling of $\xi_{L}/L$ at $x$=0.125\label{fig:sc125}}

\end{figure}

The scaling equation (\ref{eq:xi scaling}) is expected to be satisfied
only in close proximity of $T_{g}$. To better describe the data at
a larger distance from the critical point, Campbell \textit{et al.}
proposed a heuristic extended scaling scheme (ESS)\cite{CampbellExtendScal}
for $\xi_{L}/L$ of the form:\begin{equation}
\xi_{L}/L=\tilde{Y}\left(\left(TL\right)^{1/\nu}\left(1-\frac{T_{g}}{T}\right)\right),\label{eq:extended scaling}\end{equation}
and showed the improvement of the accuracy it provides in the case
of a 2D Ising ferromagnet. Based on the assumption of a symmetric
interaction distribution, $P(J_{ij})$, they proposed, and tested
numerically, an alternative scaling formula for the Ising EA spin
glass, where $\nicefrac{T_{g}}{T}$ in Eq. (\ref{eq:extended scaling})
is replaced with $\left(\nicefrac{T_{g}}{T}\right)^{2}$. Here we
use the scaling formula of Eq. (\ref{eq:extended scaling}) because
the bond distribution in the case of diluted dipoles is not symmetric.
In a recent MC simulation of a diluted dipolar Ising SG, an ESS as
given by (\ref{eq:extended scaling}) was also found to describe the
scaling of $\xi_{L}/L$ better than if $\left(\nicefrac{T_{g}}{T}\right)^{2}$
was used.\cite{Tam_LiHo}

We fit our data for $\xi_{L}/L$ to the scaling function (\ref{eq:extended scaling})
over the whole simulated temperature range, shown in Table I. The
scaling function, $\tilde{Y}$, is approximated with a 6th order polynomial,
\begin{equation}
F(z)=\sum_{m=0}^{6}a_{m}z^{m},\label{eq:F(z)}\end{equation}
 where $z=\left(TL\right)^{1/\nu}(1-\nicefrac{T_{g}}{T})$. We define
the penalty function,\begin{equation}
D=\sum_{{\rm MC\, data}}\left(F(z)L/\xi_{L}-1\right)^{2},\label{eq:D (penalty  function)}\end{equation}
 that is minimized with respect to the parameters $\left\{ a_{m}\right\} $,
$T_{c}$ and $\nu$. We obtain the values of the critical exponent
$\nu=$1.16, $\nu=$1.09, and transition temperatures $T_{g}$=0.074,
$T_{g}$=0.12 for $x$=0.0625 and $x$=0.125, respectively. The scaling
collapse of the simulation data is shown in Fig. \ref{fig:sc625}
and Fig. \ref{fig:sc125}. %
\begin{figure}
\includegraphics[scale=0.5]{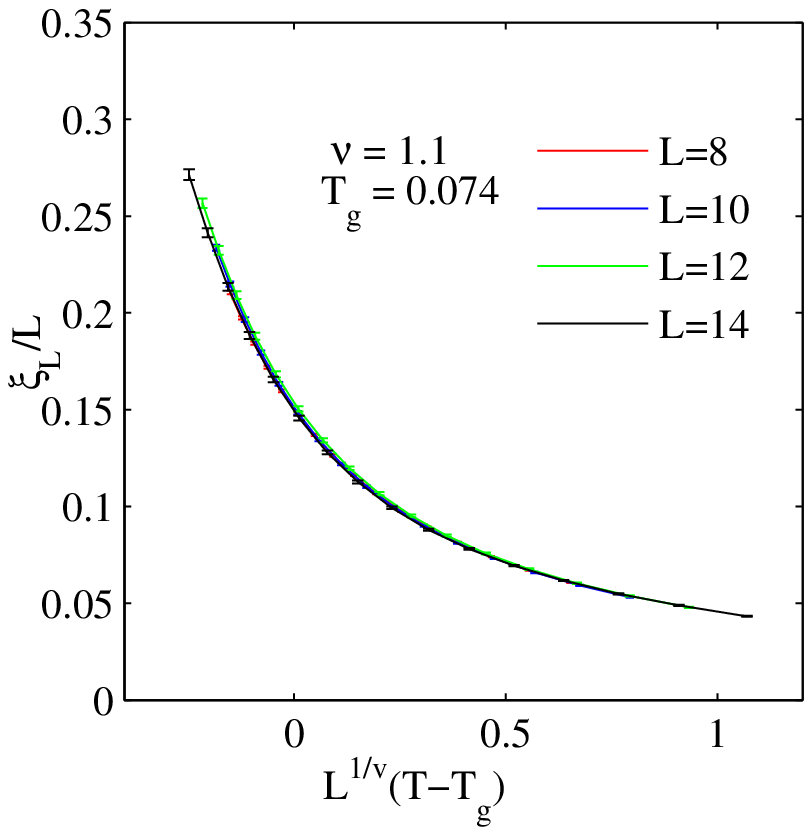}\includegraphics[scale=0.5]{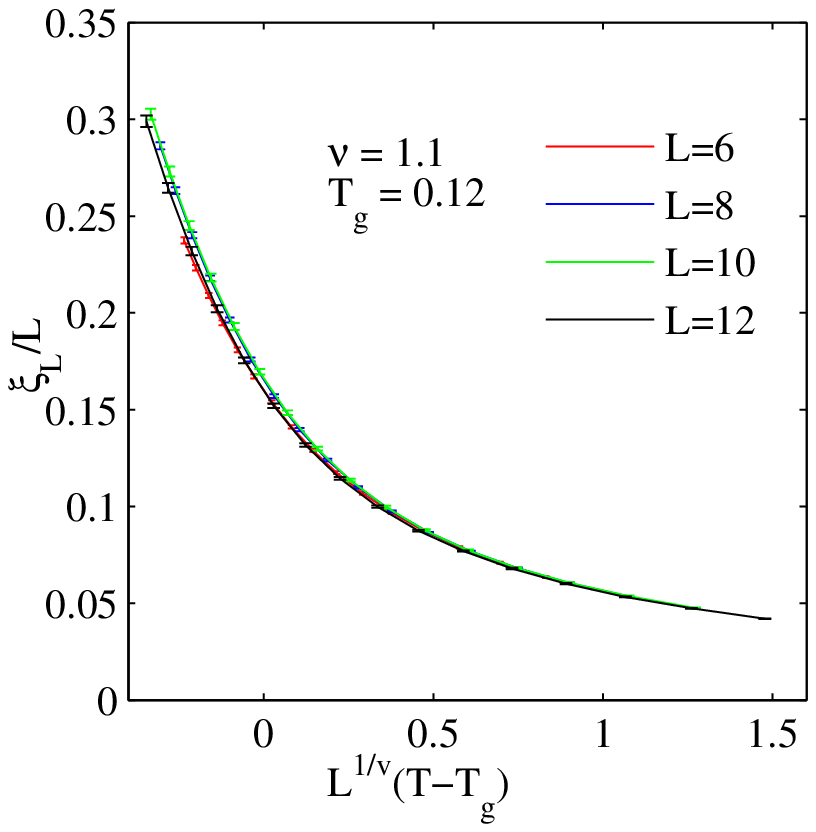}

\caption{(color online).
Conventional scaling of $\xi_{L}/L$ with $L^{1/\nu}\left(T-T_{g}\right)$;
$x$=0.0625 (left) and $x$=0.125 (right).\label{fig:sc}}

\end{figure}

\begin{figure}
\includegraphics[scale=0.5]{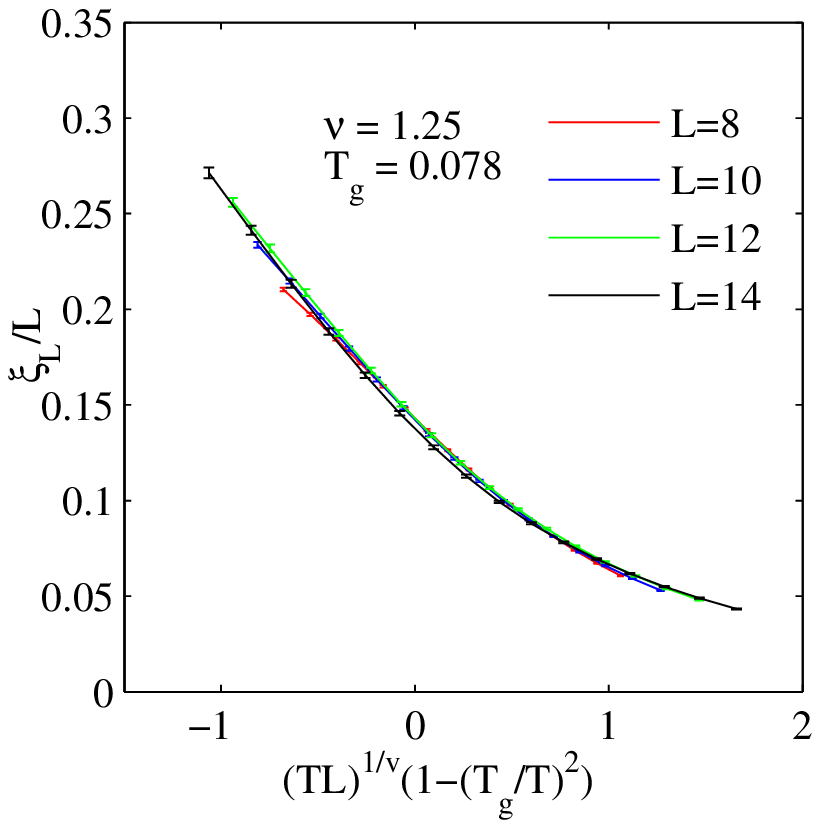}\includegraphics[scale=0.5]{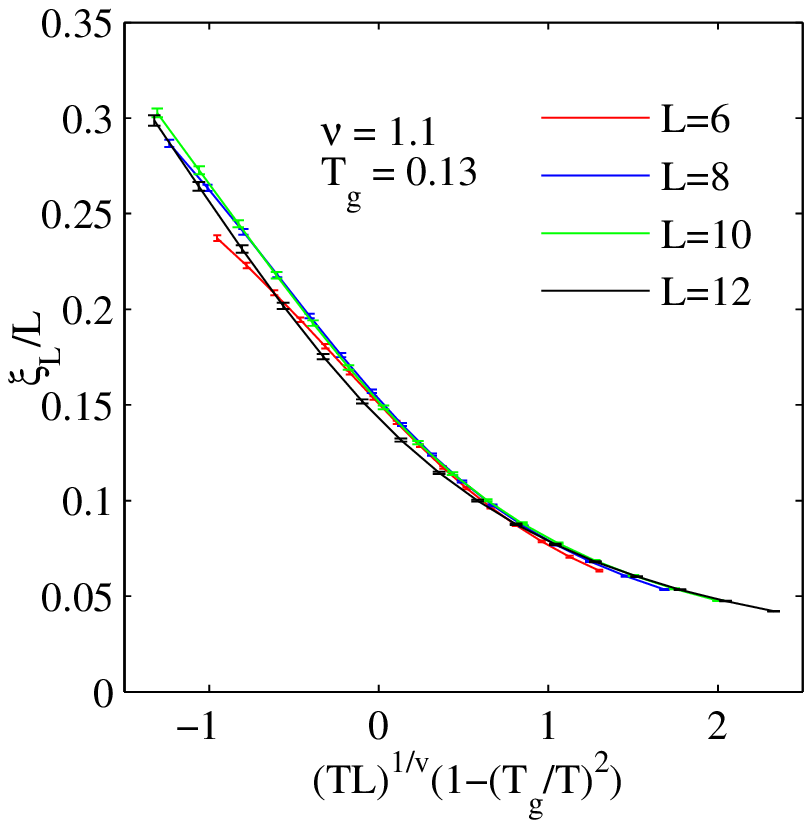}

\caption{(color online).
Extended scaling of $\xi_{L}/L$ with $\left(TL\right)^{1/\nu}\left(1-\left(\nicefrac{T_{g}}{T}\right)^{2}\right)$;
$x$=0.0625 (left) and $x$=0.125 (right). \label{fig:ex2sc}}

\end{figure}

In Figs. \ref{fig:sc} and \ref{fig:ex2sc}, just for comparison,
we present the results of the fitting to the conventional formula
(\ref{eq:xi scaling}) and ESS with $\left(\nicefrac{T_{g}}{T}\right)^{2}$.
The fitting to the conventional formula (\ref{eq:xi scaling}), shown
in Fig. \ref{fig:sc}, gives quite similar results to the ESS of Eq.
(\ref{eq:extended scaling}). Apparently, the inaccuracy due to the
small system sizes studied is larger here than the correction made
by replacing Eq. (\ref{eq:xi scaling}) with Eq. (\ref{eq:extended scaling}). 

In the case of fitting to Eq. (\ref{eq:extended scaling}) with $\left(\nicefrac{T_{g}}{T}\right)^{2}$,
we obtain a visibly worse data collapse than when $\nicefrac{T_{g}}{T}$
is used; the result of such a fit is shown in Fig. \ref{fig:ex2sc}.

\begin{figure}[h]
\includegraphics[width=0.8\columnwidth]{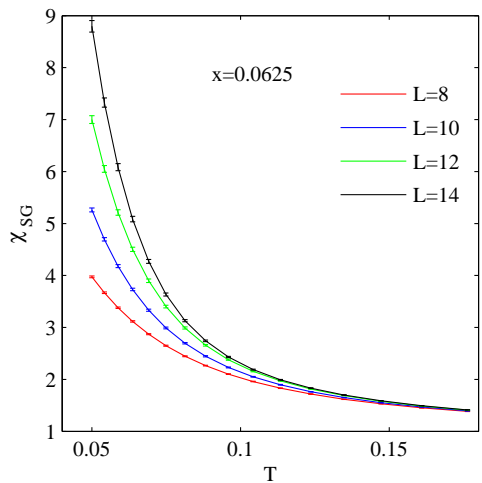}

\caption{(color online).
SG susceptibility, $x$=0.0625\label{fig:chi625}}

\end{figure}
\begin{figure}[h]
\includegraphics[width=0.8\columnwidth]{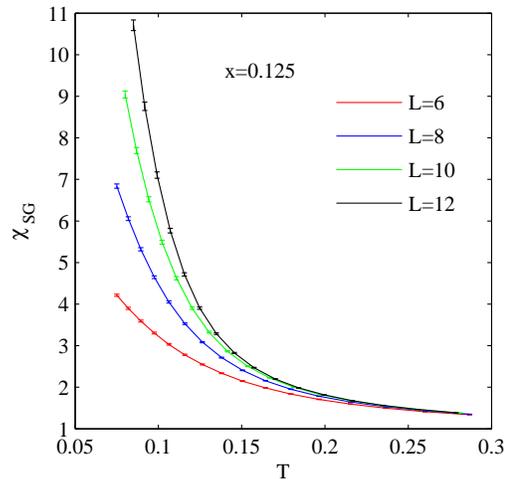}

\caption{(color online).
SG susceptibility, $x$=0.125\label{fig:chi125}}

\end{figure}
We plot in Figs. \ref{fig:chi625} and \ref{fig:chi125} the SG susceptibility
for $\bm{k}=0$, $\chi_{{\rm SG}}(0)$, of Eq. (\ref{eq:chi}) for
$x$=0.0625 and $x$=0.125, respectively. %
\begin{figure}[h]
 \includegraphics[width=0.8\columnwidth]{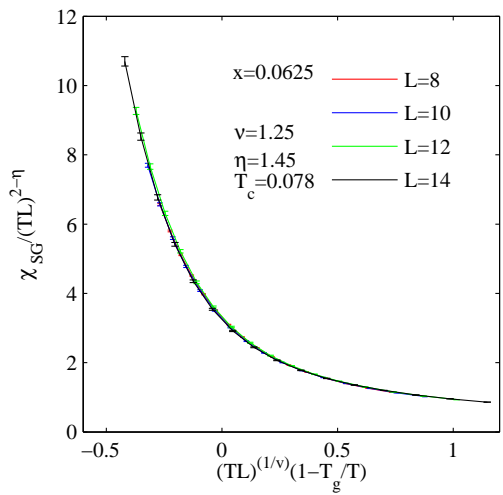}

\caption{(color online).
SG susceptibility scaling, $x$=0.0625\label{fig:chiexsc625}}

\end{figure}
\begin{figure}[h]
 \includegraphics[width=0.8\columnwidth]{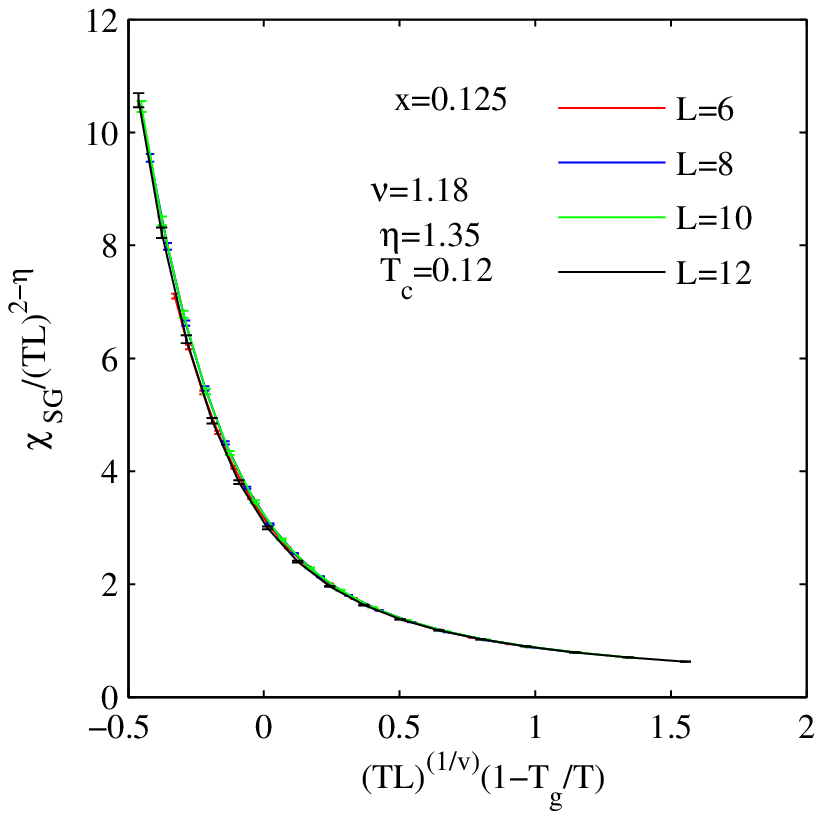}

\caption{(color online).
SG susceptibility scaling, $x$=0.125\label{fig:chiexsc125}}

\end{figure}

The SG susceptibility is expected to scale according to the ESS formula\cite{CampbellExtendScal}\begin{equation}
\chi_{{\rm SG}}=\left(TL\right)^{2-\eta}\tilde{Z}\left(\left(TL\right)^{1/\nu}\left(1-\frac{T_{g}}{T}\right)\right),\label{eq:chi extended scaling}\end{equation}
 We performed a fit following a procedure similar to the method used
for the scaling fit of $\xi_{L}/L$ described in Eqs. (\ref{eq:F(z)})
and (\ref{eq:D (penalty  function)}). For $x$=0.0625 we obtain $T_{g}$=
0.078, $\nu$=1.25 and $\eta$=1.45. For $x$=0.125 we get $T_{g}$=
0.12, $\nu$=1.18 and $\eta$=1.35. The critical temperatures are
consistent with those obtained from FSS of $\xi_{L}/L$. The values
of the critical exponent $\nu$ obtained here are slightly larger than
$\nu$ obtained from the scaling of $\xi_{L}/L$. The scaling collapse
of $\chi_{\textrm{SG}}$ is plotted in Fig. \ref{fig:chiexsc625}
and \ref{fig:chiexsc125} for $x$=0.0625 and $x$=0.125, respectively.%
{}

In the dipolar Hamiltonian (\ref{eq:H}) off-diagonal terms, that
couple different vector components of the dipolar moment, are present.
The off-diagonal terms destroy the rotational (O(3)) symmetry in an
otherwise isotropic vector spin system, and only a Z$_{2}$ symmetry
remains. It was suggested that such spatially disordered dipolar systems
belong to the Ising universality class.\cite{Bray_Moore,See_Gingras_PRL}
Due to spatial disorder, the couplings, including the off-diagonal
terms, are random, and the distribution of local freezing direction
in the SG phase remains uniform, unlike in a system with a global
single-ion anisotropy (e. g. $-DS_{z}^{2}$ term in the Hamiltonian).
With the uniform distribution of local freezing directions, a system
is said to have a statistical rotational symmetry.\cite{See_Gingras_PRL}

The values of the critical exponents found in this work do not agree
with either those from simulations of short-range Ising SG, $\nu=2.45$,
$\eta=-0.375$,\cite{Hasenbusch_Ising_SG} nor Heisenberg SG, $\nu=1.49$,
$\eta=-0.19$.\cite{Fernandez_Young} It is possible that our exponent
$\nu$ is consistent with $\nu=1.3$ and $\nu=0.95$ obtained for
Ising diluted dipolar SG in Ref.~[\onlinecite{Tam_LiHo}] and Ref.~[\onlinecite{Alonso_arXiv}],
respectively. Extracting SG critical exponents from simulations is
difficult. Critical exponents for the Ising SG have been discussed
for a long time\cite{Ogielski_Morgenstern_Ising_SG,Bhatt_Young_Ising_SG,Ballesteros_Ising_SG,katzgraber:224432,Hasenbusch_Ising_SG,See_Hasenbusch_Ising_SG},
and proposed values were changing much with progress in development
of simulation algorithms and computer hardware. Similarly to the early
simulations of the Ising SG,\cite{Ogielski_Morgenstern_Ising_SG,Bhatt_Young_Ising_SG}
our data very likely suffer from large scaling corrections. As our
system sizes are small, one may want to compare our exponent $\nu$
with the results of simulations of the Ising SG performed for small
system sizes, e.g. these in Ref.~[\onlinecite{ Ogielski_Morgenstern_Ising_SG}]
($\nu=1.2$) or Ref.~[\onlinecite{Bhatt_Young_Ising_SG}] ($\nu=1.3$).
There is a fair agreement in $\nu$ but not in $\eta$. The value
of the exponent $\eta$, from simulation\cite{Ballesteros_Ising_SG,katzgraber:224432,Hasenbusch_Ising_SG,Fernandez_Young}
and experiments\cite{Gunnarsson_FeMnTiO3,Nair_EuBaMnO,Fischer} on
many different materials, for both Ising and Heisenberg SG is a small
number, either positive or negative but not exceeding 0.5 in absolute
value. Surprisingly, the value of $\eta$ we obtain for the diluted
dipolar Heisenberg SG, $\eta=~$1.4, is much larger.

\begin{figure}
\includegraphics[width=0.8\columnwidth]{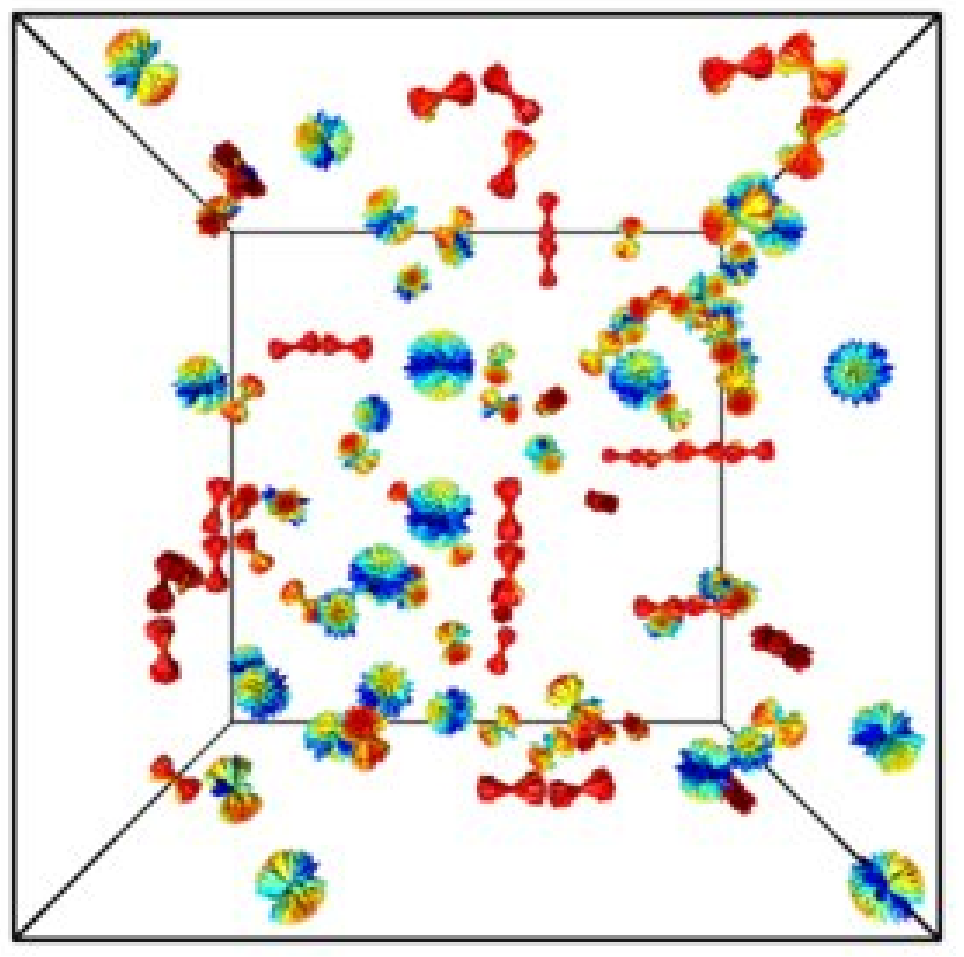}\includegraphics{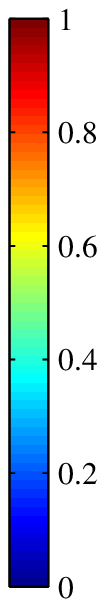}

\caption{(color online). Snapshot of 200 equilibrated independently spin configurations
for $L$=12, at $T$=0.05 and $x$=0.0625. The alignment, i.e the
scalar product, of the spins with the local freezing axes is indicated
by the colors of arrows.\label{fig:Snapshot}}

\end{figure}
Having discussed the question of universality class and commented
on the expectation that, for diluted $n$=3 component dipoles, it
should be Ising, it is interesting to ask whether such Ising structure
is explicitly physically manifest in the low temperature regime of
the systems studied above. We show in Fig. \ref{fig:Snapshot} a number
of super imposed snapshots of the spin configurations for one disorder
realization, in the low temperature phase, at $T$=0.05, and dipole
concentration $x$=0.0625. The image contains the spin configurations
of 200 replicas of the same disorder, each equilibrated independently,
starting from different random initial configuration. The system size
is $L$=12, which, at the concentration $x$=0.0625, gives $N$=108
ions. The parameters of the simulations, i.e. temperatures and number
of equilibration sweeps, are given in Table \ref{tab:Parameters}.
In the case of isotropic Heisenberg models, a low temperature phase
has O(3) rotational symmetry, and one expects the spin directions
in replicas of the same disorder as explained above to be uniformly
distributed. Here, due to the anisotropic character of the dipolar
interactions, a subset of the dipoles is characterized by a unique
Ising local freezing direction. It is indicated by the fact that in
the snapshots some dipoles have a strong tendency to point along a
particular local \textit{random} direction, i.e the arrows can be
enclosed by a circular conical surface with a small opening angle.
Such inhomogeneous {}``random Ising structures'' have also been
observed in a model of diluted two-component 2D quadrupoles. For clarity,
the alignment of spins with the local freezing directions, which is
measured as an absolute value of the scalar product of a spin and
the local freezing direction, is indicated by the color of the arrows.
The local freezing direction vector is computed by summing all the
spin vectors 
at a given site for the 200 disorder realizations
in the following way. Starting from the second element
in the sum, it is checked if adding another vector to the existing
sum will increase or decrease the magnitude of the new sum. If adding
the new element is to decrease the magnitude of the sum, the spin
vector is added with a minus sign, such that the magnitude of the
sum always increases. In this way we obtain a vector that is
pointing along the local freezing axis.
Not all the sites are characterised by a local freezing direction.
The arrows on the sites that do not have a local freezing direction
create spherical structures. These dipoles have freedom to point in
any direction in the low temperature phase. That means that these
dipoles are strongly frustrated and decoupled from the other dipoles.
It is interesting to note that this behavior resembles the presence
of {}``protected degrees of freedom'' observed in gadolinium gallium
garnet (GGG).\cite{Ghosh_GGG} The sites with a local freezing direction,
in Fig. \ref{fig:Snapshot}, seem to form small clusters. Possibility
of ferromagnetic spin blocking was mentioned earlier as a potential
explanation of a large shift of the correlation length crossing points
for the largest system sizes relative to the crossing points for the
smaller sizes. Formation of ferromagnetic spin blocks is also suggested
by nonzero, but decreasing with system size, finite-size magnetization
(see Appendix \ref{sec:Magnetization}). 

\begin{figure}
\includegraphics[width=0.5\columnwidth]{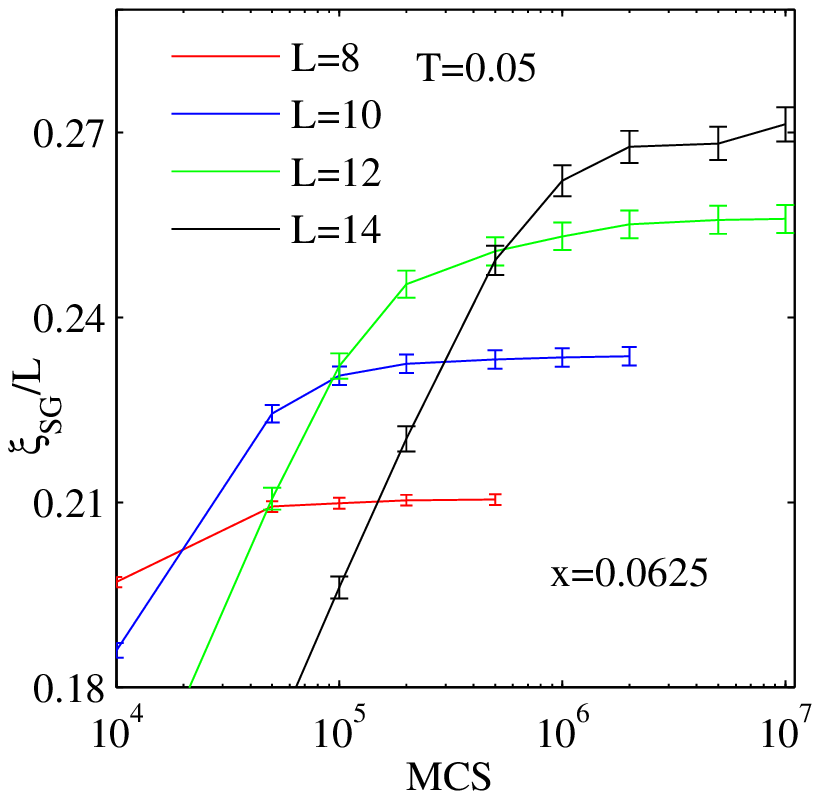}\includegraphics[width=0.5\columnwidth]{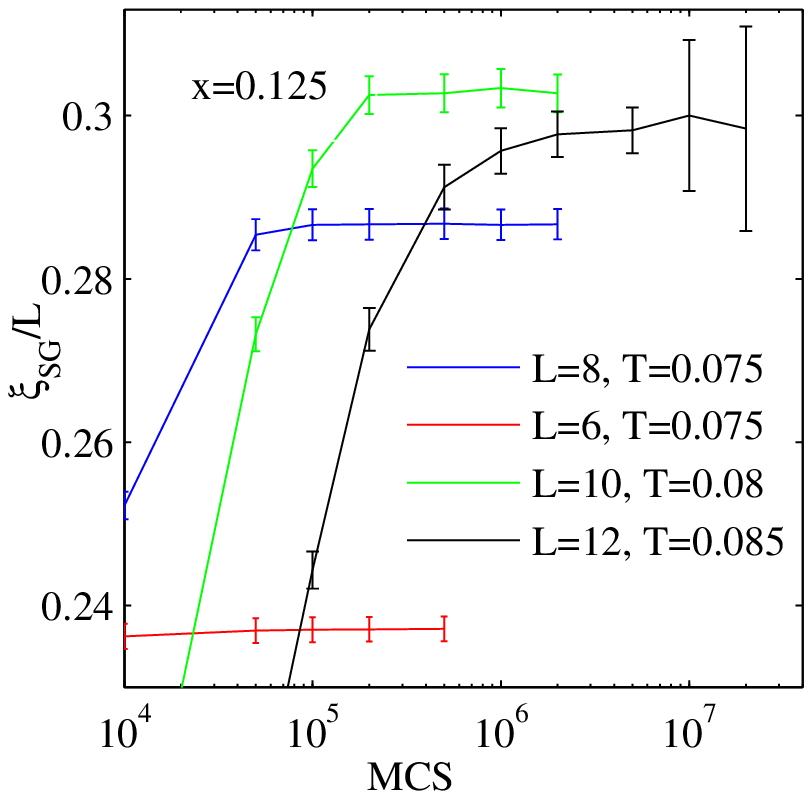}

\caption{(color online).
Equilibration $x$=0.0625 (left) and $x$=0.125 (right).\label{fig:Equlibration}}

\end{figure}

In the simulations of a SG system, it is of paramount importance to
ensure equilibration of the system before the statistics for the measured
observables is collected. As the quantity of foremost interest here
is the correlation length, we assume that the system is equilibrated
when the correlation length reaches a stationary state. We plot in
Fig. \ref{fig:Equlibration} $\xi_{L}/L$ vs the number of the equilibration
steps performed before the measurement was taken. The number of necessary
equilibration steps increases very rapidly with the system size and,
because of this, we were only able to equilibrate system sizes up
to about 200 dipoles.%
We observe that the long-range dipolar Heisenberg SG takes longer
time to equilibrate than the short-range 3D EA Heisenberg SG model.\cite{Viet_Kawamura,Fernandez_Young}
A similar fact has been observed in the case of dipolar Ising SG.\cite{Tam_LiHo}

\section{Summary\label{sec:Summary}}

In conclusion, we have studied the spin-glass (SG) transition in a
diluted dipolar Heisenberg model. From an analysis of the finite-size
scaling of the SG correlation length, $\xi_{L}$, we found an indication
of a SG transition at a temperature $T_{g}$=0.074, $T_{g}$=0.12,
and critical exponent $\nu=$1.16, $\nu=$1.09 for dipole concentrations
$x$=0.0625 and $x$=0.125, respectively. From finite-size scaling
of the  SG susceptibility, $\chi_{{\rm SG}}$, we obtained $T_{g}$= 0.078,
$\nu$=1.25, $\eta$=1.45, and $T_{g}$= 0.12, $\nu$=1.18, $\eta$=1.35
for $x$=0.0625 and $x$=0.125, respectively. As in the isotropic
Heisenberg SG, the Binder ratios, $U_{L}$, do not exhibit a crossing
for different system sizes. 

Our data support the scenario of ferromagnetic spin blocking. Short-range
ferromagnetic correlations are indicated by a relatively large finite-size
magnetization. Such short-range correlations would also explain unusual
behavior of the SG correlation length $\xi_{L}$. The crossing points
of the $\xi_{L}/L$ vs $T$ plots for the largest system sizes is
shifted to much lower temperatures from the crossing points for the
smaller system sizes. It may be caused by reaching a system size that
is larger than the length scale of ferromagnetic clustering. Some
indication of formation of frozen spin clusters can be also found
from inspection of spin configuration snapshots. 

The long-range interactions, and hence the large number of interacting
spin pairs, give rise to a larger level of random frustration than
in short-range (nearest neighbor) SG. Diluted dipolar SG seems to
be more difficult to equilibrate than nearest-neighbor models. For
example, we performed $10{}^{7}$Monte Carlo sweeps to equilibrate
a system of around 200 dipoles. To compare, with the case of the Heisenberg
Edwards-Anderson spin glass, around $10{}^{7}$ Monte Carlo sweeps,
with both overrelaxation and heatbath sweeps counted as a Monte Carlo
sweep, were used to equilibrate a system of 32,768 spins.\cite{Viet_Kawamura_arXiv}
In simulations of the Ising Edwards-Anderson spin glass around $6.5\cdot10{}^{6}$
Monte Carlo sweeps were used to equilibrate a system of 8000 spins.\cite{Ballesteros_Ising_SG}
Further progress in exploring the freezing in Ising\cite{Tam_LiHo,Alonso_arXiv}
and Heisenberg (this work) dipolar spin glasses will necessitate more
sophisticated methods. We hope that our present work motivate such
developments.
\begin{acknowledgments}
We thank Ka-Ming Tam and Paul McClarty for useful discussions. This
work was funded by NSERC, the CRC Program (M. G., Tier 1). The calculations
were made possible by dedicated resource allocation and the facilities
of the Shared Hierarchical Academic Research Computing Network (SHARCNET:www.sharcnet.ca).
\end{acknowledgments}
\appendix

\section{Magnetization and staggered magnetization\label{sec:Magnetization}}

As system of dipoles placed on the fully occupied SC lattice, or when
the fraction of vacant sites is sufficiently low, orders antiferromagnetically.\cite{Luttinger_Tisza,Romano}
To rule out the presence of a long-range order we calculate the magnetization
and staggered magnetization. The thermal and disorder averaged magnitude
of magnetization is defined as\begin{equation}
M=\left[\left\langle \left|\frac{1}{N}\sum_{i=1}^{N}\bm{S}_{i}\right|\right\rangle \right],\end{equation}
where $\left\langle \ldots\right\rangle $ denotes thermal averaging
and $\left[\ldots\right]$ is a disorder average. 

The antiferromagnetic ground state (GS) of a system of dipoles on
fully occupied SC lattice is described by a spin vector with the following
components:\cite{Luttinger_Tisza,Fernandez_antiferromagnet}\begin{equation}
\begin{array}{c}
S_{i}^{x}=\tau_{i}^{x}\sin\theta\cos\phi,\\
S_{i}^{y}=\tau_{i}^{y}\sin\theta\sin\phi\\
S_{i}^{z}=\tau_{i}^{z}\cos\theta,\end{array},\label{eq:SC dipole GS}\end{equation}
Such GS has two global rotational degrees of freedom: polar angle,
$\theta$, and azimuthal angle, $\phi$. The sublattice and direction
indexing vector, $\bm{\tau}_{i}\equiv[\tau_{i}^{x},\tau_{i}^{y},\tau_{i}^{z}${]},
is given by \begin{equation}
\bm{\tau}_{i}=[(-1)^{r_{i}^{y}+r_{i}^{z}},(-1)^{r_{i}^{x}+r_{i}^{z}},(-1)^{r_{i}^{x}+r_{i}^{y}}],\label{eq:tau}\end{equation}
where $\bm{r}_{i}$ is the position of site $i$, measured in units
of lattice constant, and its vector components, $r_{i}^{x}$, $r_{i}^{y}$
and $r_{i}^{y}$, on SC lattice, are all integers. The staggered magnetization,
which is indicating ordering described by Eqs. (\ref{eq:SC dipole GS}),
using sublattice and direction indexing vector $\bm{\tau}_{i}$ of
Eq. (\ref{eq:tau}), is given by \begin{equation}
M_{{\rm stag}}=\left[\left\langle \left|\frac{1}{N}\sum_{i=1}^{N}\bm{S}_{i}\cdot\bm{\tau}_{i}\right|\right\rangle \right].\end{equation}

\begin{figure}
\includegraphics[width=0.5\columnwidth]{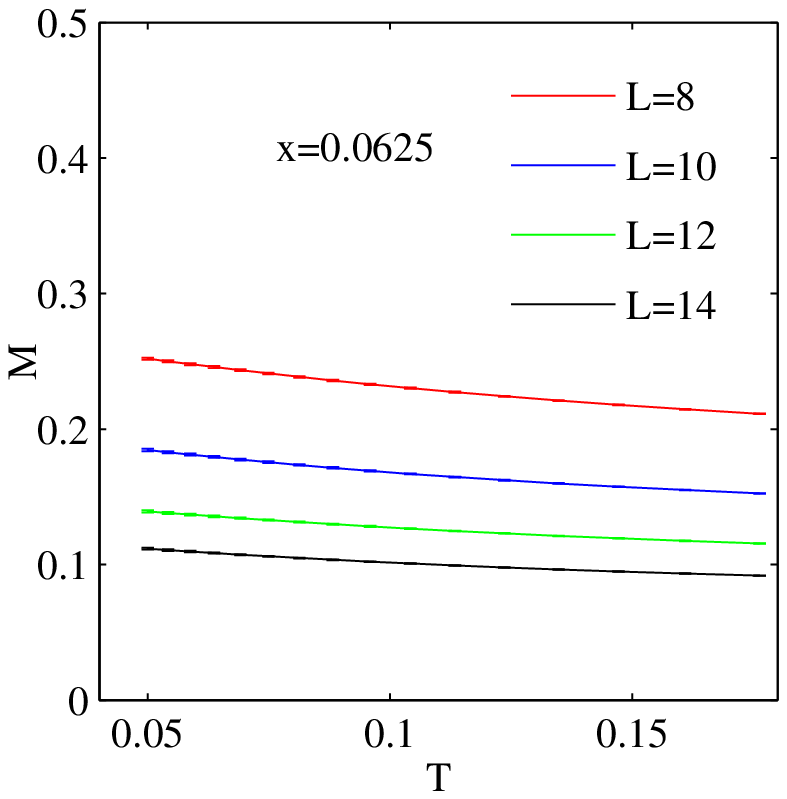}\includegraphics[width=0.5\columnwidth]{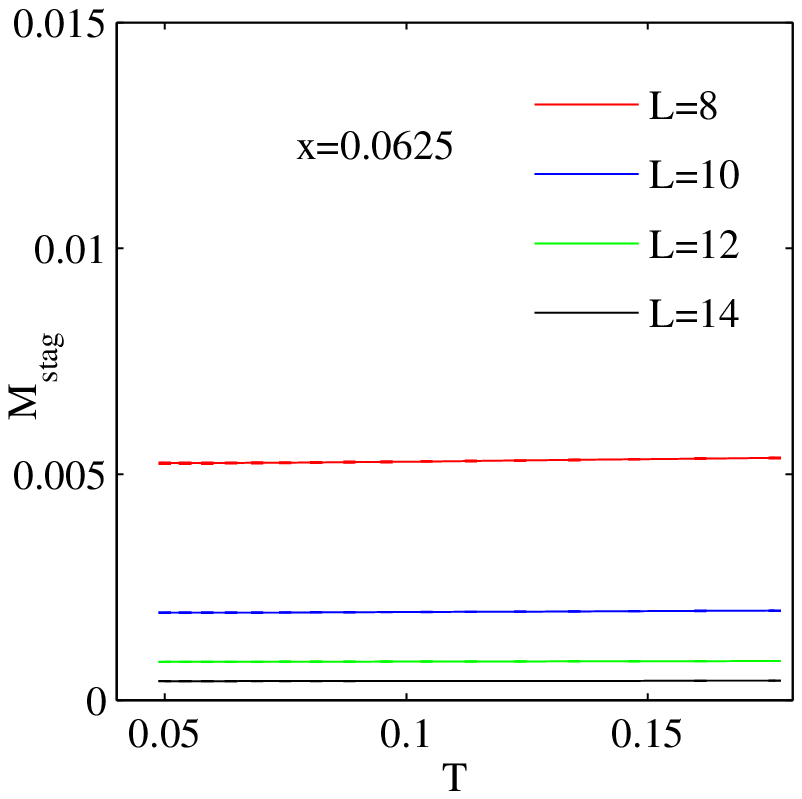}

\caption{(color online).
Magnetization, $M$, and staggered magnetization, $M_{{\rm stag}}$,
$x$=0.0625.\label{fig:Magnetization 0.0625}}

\end{figure}

\begin{figure}

\includegraphics[width=0.5\columnwidth]{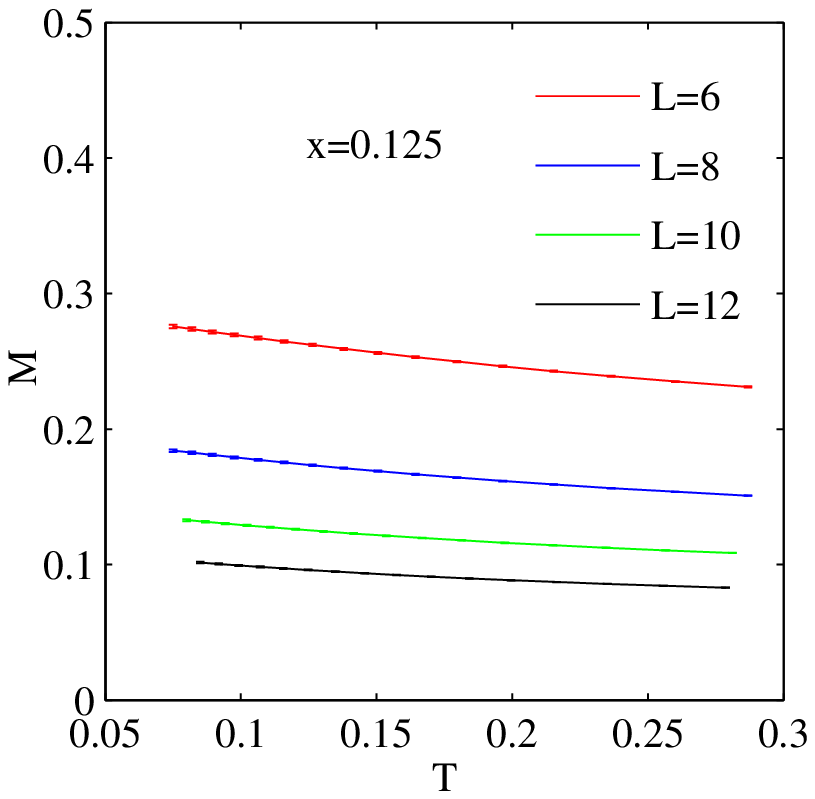}\includegraphics[width=0.5\columnwidth]{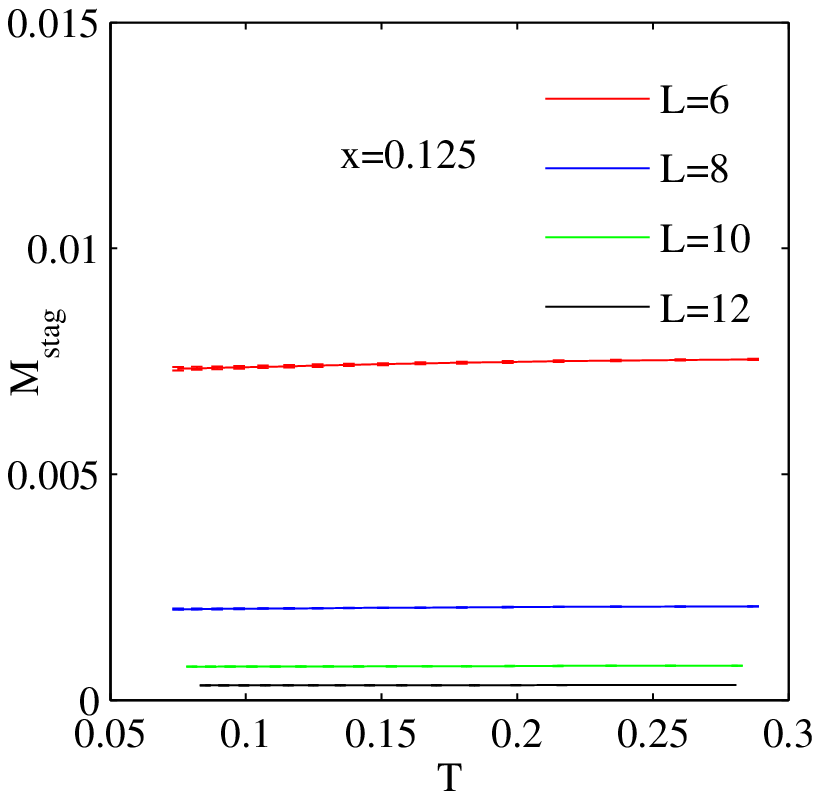}

\caption{(color online).
Magnetization, $M$, and staggered magnetization, $M_{{\rm stag}}$,
$x$=0.125.\label{fig:Magnetization 0.125}}

\end{figure}
In Figs. \ref{fig:Magnetization 0.0625} and \ref{fig:Magnetization 0.125}
we plot the magnetization, $M$, and the staggered magnetization,
$M_{{\rm stag}}$, for $x$=0.0625 and $x$=0.125, respectively. $M$
has a small value that decreases with system size, $L$. This indicates
that nonzero magnetization is just a finite-size effect and not an
indication of long-range order. Furthermore, $M$ remains constant
at all temperatures and does not increase below $T_{g}$. $M_{{\rm stag}}$,
similarly to $M$, decreases with increasing system size, $L$, and
there are no features indicating ordering transition. The fairy large
magnetization indicates that the finite-size effects are large, and
thus the scaling corrections are expected to be large. The staggered
magnetization, $M_{{\rm stag}}$, is smaller than the magnetization,
$M$. Relatively large magnetization can indicate formation of short-range
ferromagnetic blocks. Ferromagnetic spin blocking has been observed
in experimental studies of diluted dipolar Heisenberg SG systems Eu$_{x}$Sr$_{1-x}$S\,\cite{Tholence_EuSrS_dSG,Kotzler_EuSrS_dSG}
and Eu$_{x}$Ca$_{1-x}$B$_{6}$.\cite{Wigger_EuCaB}

\section{Periodic boundary conditions and self-interaction \label{sec:Self-interaction}}

We consider a dipolar Hamiltonian of the form

\begin{equation}
\mathcal{H}=\frac{1}{2}\epsilon_{d}\sum_{i,j,\mu,\nu}\frac{\delta^{\mu\nu}r_{ij}^{2}-3r_{ij}^{\mu}r_{ij}^{\nu}}{r_{ij}^{5}}S^{\mu}(\bm{r}_{i})S^{\nu}(\bm{r}_{j}),\end{equation}
 where $\mu$ and $\nu$ are vector components, $\mu$,$\nu$=$x$,
$y$ or $z$. $\mathcal{H}$ can be written as

\begin{equation}
\mathcal{H}=\frac{1}{2}\epsilon_{d}\sum_{i,j,\mu.\nu}\mathcal{L}_{ij}^{\mu\nu}S_{i}^{\mu}S_{j}^{\nu},\end{equation}
 or shorter

\begin{equation}
\mathcal{H}=\frac{1}{2}\epsilon_{d}\sum_{i,j}\bm{S}_{i}\hat{\mathcal{L}}_{ij}\bm{S}_{j},\end{equation}
 where \begin{equation}
\mathcal{L}_{ij}=\mathcal{L}(\bm{r}_{ij})=\frac{\delta_{ij}\left|\bm{r}_{ij}\right|^{2}-3r_{ij}^{\mu}r_{ij}^{\nu}}{\left|\bm{r}_{ij}\right|^{5}}.\end{equation}
 To impose periodic boundary conditions, we replace the interaction
matrix, $\mathcal{L}_{ij}$, with\begin{equation}
L_{ij}^{\mu\nu}=\left.\sum_{\bm{n}}\right.'\frac{\delta_{ij}\left|\bm{r}_{ij}+\bm{n}\right|^{2}-3(\bm{r}_{ij}+\bm{n})^{\mu}(\bm{r}_{ij}+\bm{n})^{\nu}}{\left|\bm{r}_{ij}+\bm{n}\right|^{5}},\label{eq:Lij}\end{equation}
 where $\bm{n}=kL\hat{x}+lL\hat{y}+mL\hat{z}$; $k$, $l$, $m$ are
integers and $\hat{x}$, $\hat{y}$ and $\hat{z}$ are unit vectors.
$L$ is the linear dimension of the cubic simulation box in units
of $a$, the linear size of the cubic unit cell. $\left.\sum_{\bm{n}}\right.'$
means that the summation does not include the $\bm{n}$=0 term for
$i=j$, where $\bm{r}_{ij}=0$. One must be aware of the presence
of the ($\bm{n}$$\neq0$) self-interaction term\begin{equation}
L_{ii}^{\mu\nu}=\sum_{\bm{n\neq0}}\frac{\delta_{ij}\left|\bm{n}\right|^{2}-3\bm{n}^{\mu}\bm{n}^{\nu}}{\left|\bm{n}\right|^{5}}.\label{eq:Lii selfinteraction}\end{equation}
 The self-interaction term describes the interaction of a dipole with
its own periodic images replicated outside the simulation box. For
a cubic simulation box it reduces to a simple form $L_{ii}^{\mu\nu}=L_{ii}\delta_{\mu\nu}$
. To show that the off-diagonal terms are zero, for $\mu\neq\nu$
we write\begin{equation}
L_{ii}^{\mu\nu}=-3\sum_{\bm{n\neq0,\,\bm{n}^{\mu}>0}}\frac{\bm{n}^{\mu}\bm{n}^{\nu}+\left(-\bm{n}^{\mu}\right)\bm{n}^{\nu}}{\left|\bm{n}\right|^{5}}=0.\end{equation}
And further, in cubic symmetry, all three directions $\hat{x}$, $\hat{y}$
and $\hat{z}$ are equivalent; hence, $L_{ii}^{xx}=L_{ii}^{yy}=L_{ii}^{zz}.$

\section{Ewald summation\label{sec:Ewald}}

We wish to calculate the lattice sum

\begin{equation}
L_{ij}^{\mu\nu}=\left.\sum_{\bm{n}}\right.'\frac{\delta_{ij}\left|\bm{r}_{ij}+\bm{n}\right|^{2}-3(\bm{r}_{ij}+\bm{n})^{\mu}(\bm{r}_{ij}+\bm{n})^{\nu}}{\left|\bm{r}_{ij}+\bm{n}\right|^{5}}.\label{eq:Ewald sum 1}\end{equation}
 Again, the prime symbol in the summation sign means that for $i=j$
the sum does not include the $\bm{n}$=0 term. Noting that

\begin{equation}
-\nabla_{\mu}\nabla_{\nu}\frac{1}{\bm{r}}=\frac{\delta_{\mu\nu}r^{2}-3r_{\mu}r_{\nu}}{r^{5}},\end{equation}
 we write

\begin{equation}
L_{ij}^{\mu\nu}=-\nabla_{\mu}\nabla_{\nu}\left.\sum_{\bm{n}}\right.'\frac{1}{\left|\bm{r}_{ij}+\bm{n}\right|};\label{eq:Lij deriv sum}\end{equation}
 hence, we may calculate the lattice summation for a Coulomb potential
and obtain the sums for dipolar interactions by taking derivatives
afterwards.

The infinite sum (\ref{eq:Lij deriv sum}) is conditionally convergent,
meaning that the result depends on the asymptotic order of summation.
The Coulomb or dipolar potential is slowly decaying at large distances;
hence, with direct summation, it converges slowly. To alleviate these
problems, the summation is performed using the method introduced by
Ewald.\cite{Ewald,Dove,Leeuw_Perram,Holm_Wang} In the Ewald technique,
we separate the summation into two rapidly convergent sums: one performed
in the direct (real) space and the other sum performed in the reciprocal
space. Here we show only a simplified derivation, rigorous mathematical
proofs and detailed discussions can be found in Ref.~[\onlinecite{Leeuw_Perram}].

Using the relation \begin{equation}
\frac{1}{r}=\frac{2}{\sqrt{\pi}}\intop_{0}^{\infty}e^{-r^{2}\rho^{2}}d\rho,\end{equation}
 we write

\begin{equation}
\frac{1}{r}=\frac{2}{\sqrt{\pi}}\intop_{0}^{\alpha}e^{-r^{2}\rho^{2}}d\rho+\frac{\textrm{erfc}(\alpha r)}{r},\label{eq:1/r separated}\end{equation}
 where \begin{equation}
\textrm{erfc}(x)=1-\textrm{erf}(x)=\frac{2}{\sqrt{\pi}}\int_{x}^{\infty}e^{-y^{2}}dy\end{equation}
 is the complementary error function. The second term in Eq. (\ref{eq:1/r separated}),
for large $\alpha$, is decreasing fast with increasing $r$; hence,
it converges rapidly in the summation over $\bm{n}$. The first term
falls to zero slowly with increasing $r$, but it converges rapidly
in a reciprocal space summation formulation. The splitting parameter
$\alpha$ is chosen such that both real space and reciprocal space
sums are converging equivalently rapidly. To obtain the reciprocal
space summation term we use the relation\begin{equation}
\frac{2}{\sqrt{\pi}}\sum_{\bm{n}}e^{-(\bm{r}+\bm{n})^{2}\rho^{2}}=\frac{2\pi}{L^{3}}\sum_{\bm{K}}\rho^{-3}e^{-K^{2}/4\rho^{2}}e^{i\bm{K\cdot\bm{r}}},\end{equation}
 where $\bm{K}$ are the reciprocal lattice vectors, $\bm{n}=L(k\hat{x}+l\hat{y}+m\hat{z})$;
$k$, $l$, $m$ are integers and $\hat{x}$, $\hat{y}$ and $\hat{z}$
are unit vectors. Some care must be taken to account for the particular
case of $r$=0 that corresponds to the self-interaction (\ref{eq:Lii selfinteraction}).
In that case, $n$=0 should be excluded from the summation (\ref{eq:Lij deriv sum})
and we write\begin{equation}
\frac{2}{\sqrt{\pi}}\sum_{\bm{n}\neq0}e^{-(\bm{r}+\bm{n})^{2}\rho^{2}}=\frac{2\pi}{L^{3}}\sum_{\bm{K}}\rho^{-3}e^{-K^{2}/4\rho^{2}}e^{i\bm{K\cdot\bm{r}}}-\frac{2}{\sqrt{\pi}}e^{-\bm{r}^{2}\rho^{2}}.\label{eq:r sum}\end{equation}
 Noting that

\begin{equation}
\intop_{0}^{\alpha}d\rho\,\rho^{-3}e^{-K^{2}/4\rho^{2}}=\frac{2}{K^{2}}e^{-K^{2}/4\alpha^{2}},\label{eq:r int}\end{equation}
 we can write

\begin{eqnarray}
\left.\sum_{\bm{n}}\right.'\frac{1}{\left|\bm{r}_{ij}+\bm{n}\right|} & = & \left.\sum_{\bm{n}}\right.'\frac{\textrm{erfc}(\alpha\left|\bm{r}_{ij}+\bm{n}\right|)}{\left|\bm{r}_{ij}+\bm{n}\right|}\nonumber \\
 &  & +\sum_{\bm{K}\neq0}\frac{4\pi}{L^{3}K^{2}}e^{-K^{2}/4\alpha^{2}}e^{i\bm{K\cdot\bm{r}}_{ij}}\nonumber \\
 &  & -\frac{2\alpha}{\sqrt{\pi}}\delta_{ij}.\label{eq:Coulomb sum}\end{eqnarray}
The divergent, $\bm{K}$=0 term in the reciprocal lattice summation
is omitted. %
{}

To calculate the dipolar sum in (\ref{eq:Lij deriv sum}), we need
to take derivative of expression (\ref{eq:r sum}). To start, we compute

\begin{equation}
-\nabla_{\mu}\nabla_{\nu}\frac{\textrm{erfc}(\alpha r)}{r}=\frac{\delta_{\mu\nu}B(r)r^{2}-C(r)r_{\mu}r_{\nu}}{r^{5}},\end{equation}
 where\begin{equation}
B(r)=\textrm{erfc}(r)+\frac{2\alpha r}{\sqrt{\pi}}e^{-\alpha^{2}r^{2}},\end{equation}
 and\begin{equation}
C(r)=3\textrm{erfc}(r)+\frac{2\alpha r(3+2\alpha^{2}r^{2})}{\sqrt{\pi}}e^{-\alpha^{2}r^{2}}.\end{equation}
 For the reciprocal space part we compute \[
-\nabla_{\mu}\nabla_{\nu}e^{i\bm{K\cdot\bm{r}}}=K_{\mu}K_{\nu}e^{i\bm{K\cdot\bm{r}}}.\]
To obtain the self term (the last term in Eq. \ref{eq:r sum}) we
write\begin{equation}
-\nabla_{\mu}\nabla_{\nu}e^{-\bm{r}^{2}\rho^{2}}=2\rho^{2}\left(\delta_{\mu\nu}-2\rho^{2}r_{\mu}r_{\nu}\right)e^{-r^{2}\rho^{2}},\end{equation}
and integrating (see Eq. \ref{eq:r int}) we get $-\frac{4\alpha^{3}}{3\sqrt{\pi}}\delta_{\mu\nu}\delta_{ij}$.
Finally, we have

\begin{eqnarray}
L_{ij}^{\mu\nu} & = & \left.\sum_{\bm{n}}\right.'\frac{\delta_{\mu\nu}B(r_{ij})r_{ij}^{2}-C(r_{ij})r_{ij}^{\mu}r_{ij}^{\nu}}{r^{5}}\\
 &  & +\frac{4\pi}{L^{3}}\sum_{\bm{K}\neq0}\frac{K_{\mu}K_{\nu}}{K^{2}}e^{-K^{2}/4\alpha^{2}}e^{i\bm{K\cdot\bm{r}_{ij}}}\\
 &  & -\frac{4\alpha^{3}}{3\sqrt{\pi}}\delta_{\mu\nu}\delta_{ij}.\end{eqnarray}
 Similarly to the Coulomb case (\ref{eq:Coulomb sum}), the divergent
$\bm{K}$=0 term is omitted in the reciprocal space summation.

The effect of the magnetic polarization of the surface does not vanish
in the thermodynamic limit. To model the experimental case of a spherical
sample, a direct (real space) sum (\ref{eq:Ewald sum 1}) can be computed
via summing over series of spherical shells of radius $r_{k}$, where
each shell consist of all vectors $\bm{n}$ such that $r_{k}<\left|\bm{n}\right|<r_{k+1}$.
In the Ewald method, to obtain a result equivalent to such a summation,
the surface contribution to the total energy should be included, and
it is of the form\cite{Holm_Wang} \[
U^{\textrm{(surf)}}=\frac{2\pi}{(2\epsilon'+1)L^{3}}\sum_{i,j}\bm{\mu}_{i}\cdot\bm{\mu}_{j},\]
 where $\epsilon'$ is the magnetic permeability of the surrounding
medium. In the case of a long cylindrical shape the surface term is
zero. In our simulations we set the surface term to zero and are therefore
implicitly considering a long cylindrical sample. In practice, we
set $\epsilon'=\infty$, infinite magnetic permeability outside the
considered system, the so-called {}``metallic boundary conditions'',
by analogy to the physical situation with electric, as opposed to
magnetic dipoles.

\section{Overrelaxation\label{sec:Overrelaxation}}

It has been reported that supplementing canonical Metropolis spin
updates with computationally inexpensive {}``overrelaxation'' steps
of zero energy change can substantially reduce autocorrelation times.\cite{Creutz_overrelaxation,Alonso_overrelaxation}
Unfortunately, this technique does not provide much of a performance
improvement in the case of long-range interactions and cannot be used
when periodic boundary condition are imposed on a system characterized
by dipolar interaction with non-cubic lattice symmetry.

In overrelaxation update, a new spin direction, $\bm{S}'_{i}$, is
obtained by performing a reflection of the spin at site $i$, $\bm{S}{}_{i}$,
around the local dipolar field vector, $\bm{H}_{i}$, \begin{equation}
\bm{S}'_{i}=-\bm{S}_{i}+2\frac{\bm{S}_{i}\cdot\bm{H}_{i}}{H_{i}^{2}}\bm{H}_{i}.\label{eq:overrelaxation}\end{equation}
 %
{}The local dipolar field is given by Eq. (\ref{eq:Hk}),\begin{equation}
\bm{H}_{k}=\sum_{j\neq k}\hat{L}_{kj}\bm{S}_{j},\label{eq:Hk_Ovrn}\end{equation}
 where the tensor $\hat{L}_{kj}$ stands for the dipolar interaction,
as defined in Eq. (\ref{eq:Lsum}). Using Eq. (\ref{eq:Hk_Ovrn}),
the finite-size Hamiltonian of Eq. (\ref{eq:H with L}) can be written
in the form\begin{equation}
\mathcal{H}=-\frac{1}{2}\sum_{k}\bm{S}_{k}\cdot\bm{H}_{k}-\frac{1}{2}\sum_{k}\bm{S}_{k}\hat{L}_{kk}\bm{S}_{k},\label{eq:H overrelation}\end{equation}
where the local dipolar field, $\bm{H}_{k}$, does not include the
self-term and the self term is written explicitly. Let us consider
an overrelaxation move of spin $\bm{S}_{i}$. To make the effect of
the spin move clear, we write the energy of a spin configuration before
the spin move in the form \begin{eqnarray}
E & = & -\frac{1}{2}\bm{S}_{i}\cdot\bm{H}_{i}-\frac{1}{2}\sum_{k\ne i}\bm{S}_{k}\cdot\bm{H}_{k}-\frac{1}{2}\sum_{k}\bm{S}_{k}\hat{L}_{kk}\bm{S}_{k},\label{eq:Ei}\end{eqnarray}
which is just Eq. (\ref{eq:H overrelation}) rewritten with the term
for spin $\bm{S}_{i}$ excluded from the summation and written explicitly.
After changing spin $\bm{S}_{i}$ to $\bm{S}_{i}'$, according to
Eq. (\ref{eq:overrelaxation}), we have

\begin{equation}
E'=-\frac{1}{2}\bm{S}'_{i}\cdot\bm{H}_{i}-\frac{1}{2}\sum_{k\ne i}\bm{S}_{k}\cdot\bm{H}'_{k}-\frac{1}{2}\sum_{k}\bm{S}'_{k}\hat{L}_{kk}\bm{S}'_{k},\label{eq:Ep}\end{equation}
 where $\bm{H}'_{k}$ are updated dipolar fields; $\bm{H}_{i}'=\bm{H}_{i}$
and for $k\neq i$ :\begin{equation}
\bm{H}'_{k}=\bm{H}_{k}+\hat{L}_{ki}(\bm{S}'_{i}-\bm{S}_{i}).\label{eq:Hp}\end{equation}
Combined together, Eqs. (\ref{eq:overrelaxation}), (\ref{eq:Ei}),
(\ref{eq:Ep}) and (\ref{eq:Hp}) give\begin{equation}
E'-E=\frac{1}{2}\left(\bm{S}'_{i}\hat{L}_{ii}\bm{S}'_{i}-\bm{S}_{i}\hat{L}_{ii}\bm{S}_{i}\right).\end{equation}
 The energy does not change only if $\bm{S}'_{i}\hat{L}_{ii}\bm{S}'_{i}-\bm{S}_{i}\hat{L}_{ii}\bm{S}_{i}=0$.
This is the case when, for each $\mu$,$\nu$, $L_{ii}^{\mu\nu}=0$,
or for diagonal $\hat{L}_{ii}$, $L_{ii}^{\mu\nu}=L_{ii}\delta_{\mu\nu}$,
(recalling $\left|\bm{S}_{i}'\right|=\left|\bm{S}_{i}\right|=1$),
which is satisfied in the case of cubic lattice symmetry (see Appendix
\ref{sec:Self-interaction}).

The fact that we do not use the overrelaxation method does not cause
a large decrease of efficiency in our simulation. In the case of long-range
interaction, the reflection (\ref{eq:overrelaxation}) would have
to be followed by the recalculation of dipolar field, $\bm{H}'_{k}$,
of Eq. (\ref{eq:Hp}). A similar lattice sum has to be performed in
the case of Metropolis updates. Most of the computation time is spent
on doing such lattice sums. Hence, even if it was doable, an overrelaxation
move would be practically as computationally expensive as a Metropolis
update.

\section{Heatbath algorithm\label{sec:Heatbath-algorithm}}

In the original Metropolis algorithm a random configuration update
is attempted and it is accepted with a probability that depends on
the change of energy following such configuration change. The updates
lowering the energy are always accepted while, if the energy is to
increase, the acceptance probability is\begin{equation}
P(\Delta E)=\exp(-\beta\Delta E),\end{equation}
 where $\beta=1/k_{{\rm B}}T$. The probability exponentially decreases
with an increase of the energy change, $\Delta E$. Thus, to obtain
a sufficient acceptance rate, the attempted moves have to be sufficiently
small. Usually the configuration update is chosen in such way that
the acceptance rate is close to 50\%. In many applications a better
way of performing a local spin updates is the {}``heatbath'' algorithm,\cite{Creutz_heatbath,Miyatake_heatbath,Olive_heatbath}
where the new direction of a spin is drawn from a suitable probability
distribution such that the new configuration energy is distributed
according to a Boltzmann weight. In the case of isotropic (O(3)) Heisenberg
model, the distribution of angle $\theta$ between the local dipolar
field, $\bm{H}_{i}$, and the spin vector $\bm{S}_{i}$ can be calculated
analytically.\cite{Creutz_heatbath,Miyatake_heatbath,Olive_heatbath}
In such an isotropic case, the Hamiltonian can be written as \begin{equation}
\mathcal{H}=-\frac{1}{2}\sum_{i}\bm{S}_{i}\cdot\bm{H}_{i}\label{eq:Hheatbath}\end{equation}
 where $\bm{H}_{i}$ is the interaction field and $\bm{H}_{i}$ does
not depend on spin $\bm{S}_{i}$. We did not include here the self-interaction
term because the calculation shown below is not possible with a general
self-interaction term included. The case of diagonal self-interaction
term, as in the case of cubic symmetry, will be discussed at the end
of this appendix. In the case of long-range interaction we write \begin{equation}
\bm{H}_{i}=\sum_{j\neq i}\hat{L}_{ij}\bm{S}_{j}\label{eq:HiHeatbath}\end{equation}
 It is convenient to describe spin $\bm{S}_{i}$ in polar coordinates,
$\theta$ and $\phi$, with the polar axis along the local dipolar
field, $\bm{H}_{i}$. The energy of spin $i$ in the field of other
spins is\begin{equation}
E_{i}=-\bm{S}_{i}\cdot\bm{H}_{i}=-H_{i}\cos(\theta),\label{eq:E_heatbath}\end{equation}
 where $\theta$ is the polar angle defined as the angle between $\bm{S}_{i}$
and $\bm{H}_{i}$. We wish to randomly choose $\bm{S}_{i}$ such that
the probability distribution of the energy (\ref{eq:E_heatbath})
given by Boltzmann distribution. The energy does not depend on the
azimuthal angle, $\phi$; hence, $\phi$ is randomly chosen from the
uniform distribution on the interval $[0,2\pi]$. The polar angle,
$\theta$, is chosen such that $x=\cos(\theta)$ is given by the probability
distribution\begin{equation}
P(x)=\frac{e^{\beta H_{i}x}}{\int_{-1}^{1}dx\, e^{\beta H_{i}x}}=\frac{\beta H_{i}}{2\sinh\beta H_{i}}e^{\beta H_{i}x},\label{eq:P(x)}\end{equation}
 where $\beta=1/T$. To obtain random variable $x$ drawn from distribution
(\ref{eq:P(x)}), we calculate the cumulative distribution\begin{equation}
F(x)=\int_{-1}^{x}P(x')dx'=\frac{e^{\beta H_{i}x}-e^{-\beta H_{i}}}{e^{\beta H_{i}}-e^{-\beta H_{i}}}\label{eq:F(x)}\end{equation}
 and we reverse $r\equiv F(x)$, where $r\in[0,1]$ is a uniformly
distributed random number. We obtain\cite{Creutz_heatbath,Miyatake_heatbath,Olive_heatbath}\begin{equation}
x=\frac{1}{\beta H_{i}}\ln\left[1+r\left(e^{2\beta H_{i}}-1\right)\right]-1.\label{eq:x}\end{equation}
 Having chosen $\phi$ and $\theta$, we need to compute the vector
components of the spin and rotate them to the global coordinate system.
Using the coordinates $\phi$ and $\theta$, relative to the local
molecular field, $\bm{H}_{k}$, we compute the new spin vector, $\bm{S}_{i}=(S_{x},S_{y},S_{z})$,
as follows. Let $\phi_{H}$ and $\theta_{H}$ denote azimuthal and
polar angle of vector $\bm{H}_{i}$ in global coordinates. A possible
choice of the local coordinates $\hat{x}'$, $\hat{y}'$ and $\hat{z}'$,
having $\hat{z}'$ axis along $\bm{H}_{i}$ is\begin{eqnarray}
\hat{x}' & = & \cos(\theta_{H})\cos(\phi_{H})\hat{x}+\cos(\theta_{H})\sin(\phi_{H})\hat{y}-\sin(\theta_{H})\hat{z}\nonumber \\
\hat{y}' & = & -\sin(\phi_{H})\hat{x}+\cos(\phi_{H})\hat{y}\label{eq:x'}\\
\hat{z}' & = & \sin(\theta_{H})\cos(\phi_{H})\hat{x}+\sin(\theta_{H})\sin(\phi_{H})\hat{y}+\cos(\theta_{H})\hat{z}.\nonumber \end{eqnarray}
 The new spin, $\bm{S}_{i}$, in local coordinates is\begin{equation}
\bm{S}_{i}=\sin(\theta)\cos(\phi)\hat{x}'+\sin(\theta)\sin(\phi)\hat{y}'+\cos(\theta)\hat{z}',\label{eq:SiHeatbath}\end{equation}
 and finally, combining Eq. (\ref{eq:x'}) and Eq. (\ref{eq:SiHeatbath}),
we get\begin{eqnarray}
S_{x} & = & \Theta\cos(\phi_{H})-\sin(\theta)\sin(\phi)\sin(\phi_{H}),\nonumber \\
S_{y} & = & \Theta\sin(\phi_{H})+\sin(\theta)\sin(\phi)\cos(\phi_{H}),\\
S_{z} & = & -\sin(\theta)\cos(\phi)\sin(\theta_{H})+\cos(\theta)\cos(\theta_{H}),\nonumber \end{eqnarray}
 where\begin{equation}
\Theta=\sin(\theta)\cos(\phi)\cos(\theta_{H})+\cos(\theta)\sin(\theta_{H}).\end{equation}

Hamiltonian (\ref{eq:Hheatbath}) with dipolar field (\ref{eq:HiHeatbath})
does not include a self-interaction term. In the case of dipolar interaction
with periodic boundary condition we have to include such a self-interaction
term as in the Hamiltonian of Eq. (\ref{eq:H overrelation}). For
clarity, we write this Hamiltonian again \begin{equation}
\mathcal{H}=-\frac{1}{2}\sum_{i}\left(\bm{S}_{i}\bm{H}_{i}+\bm{S}_{i}\hat{L}_{ii}\bm{S}_{i}\right).\label{eq:HSLiiS}\end{equation}
For Hamiltonian (\ref{eq:HSLiiS}), as opposed to Hamiltonian (\ref{eq:Hheatbath}),
for a general form of the matrix $\hat{L}_{ii}$, the cumulative distribution
(\ref{eq:F(x)}) cannot be integrated and reversed analytically. However,
for cubic symmetry, the self-interaction term is of the form $L_{ii}^{\mu\nu}=L_{ii}\delta_{\mu\nu}$
and Eq. (\ref{eq:HSLiiS}) reduces to\begin{equation}
\mathcal{H}=-\frac{1}{2}\sum_{i}\left(\bm{S}_{i}\bm{H}_{i}+L_{ii}\right);\label{eq:HLii}\end{equation}
 hence, it is of the form (\ref{eq:Hheatbath}), with just a constant
independent of spin configuration added, and the heatbath method can
be applied.

Although we have cubic symmetry in our system and the heatbath algorithm
could in principle be used, we decided to employ a method generally
applicable for any lattice symmetry and we did not use the heatbath
method.

\bibliography{HeisenbergSG_paper}

\begin{thebibliography}{106}
\bibitem[{foo({\natexlab{a}})}]{foot_Ayton}
\bibinfo{note}{However, the case of dense amorphous systems with Ising spins
  can remain ferromagnetic~[\onlinecite{Ayton_PRL, Ayton_PRE}].}
\bibitem[{foo({\natexlab{b}})}]{foot_ignore}
\bibinfo{note}{We also ignore week Gd$^{3+}$ single-ion anisotropy that would
  differ between materials.}
\bibitem[{foo({\natexlab{c}})}]{foot_universal}
\bibinfo{note}{We do not call the function $\tilde{X}$ universal because it
  implicitly contains a non-universal factor, as opposite to the case when a
  scaling formula is written as $U_{L}=\tilde{X}'(AL^{1/\nu}(T-T_{g}))$, with a
  non-universal constant, $A$, given explicitly and $\tilde{X}'$ being a
  universal scaling function.\cite{katzgraber:224432, Privman_scaling}}.
\bibitem[{foo({\natexlab{d}})}]{foot_smoothUL}
\bibinfo{note}{In the case of parallel tempering simulations the data are
  always smooth due to the correlations between the thermal replicas.}
\end{thebibliography}

\end{document}